\RequirePackage{rotating}
\documentclass[twocolumn, tighten]{aastex631}
\usepackage{amsmath}
\usepackage[caption=false]{subfig}
\usepackage{xfrac}
\usepackage{CJK}

\usepackage{placeins}

\shorttitle{Constraints on C/O ratio of DR Tau}
\shortauthors{Huang et al.}

\begin{document}
\begin{CJK*}{UTF8}{gbsn} 
\title{Constraints on the gas-phase C/O ratio of DR Tau's outer disk from CS, SO, and C$_2$H observations}

\correspondingauthor{Jane Huang}
\email{jane.huang@columbia.edu}
\author[0000-0001-6947-6072]{Jane Huang}
\affiliation{Department of Astronomy, Columbia University, 538 W. 120th Street, Pupin Hall, New York, NY 10027, United States of America}
\author[0000-0003-4179-6394]{Edwin A. Bergin}
\affiliation{Department of Astronomy, University of Michigan, 323 West Hall, 1085 S. University Avenue, Ann Arbor, MI 48109, United States of America}
\author[0000-0003-1837-3772]{Romane Le Gal}
\affiliation{Univ. Grenoble Alpes, CNRS, IPAG, F-38000 Grenoble, France}
\affiliation{IRAM, 300 rue de la piscine, F-38406 Saint-Martin d'H\`{e}res, France}
\author[0000-0003-2253-2270]{Sean M. Andrews} \affiliation{Center for Astrophysics \textbar\ Harvard \& Smithsonian, 60 Garden St., Cambridge, MA 02138, USA}
\author[0000-0001-7258-770X]{Jaehan Bae}
\affiliation{Department of Astronomy, University of Florida, Gainesville, FL 32611, United States of America}
\author[0000-0001-5849-577X]{Luke Keyte}
\affiliation{Department of Physics and Astronomy, University College London, Gower Street, WC1E 6BT London, United Kingdom.}
\author[0000-0002-0377-1316]{J. A. Sturm}
\affiliation{Leiden Observatory, Leiden University, P.O. Box 9513, NL-2300 RA Leiden, The Netherlands}

\begin{abstract}
Millimeter wavelength observations of Class II protoplanetary disks often display strong emission from hydrocarbons and high CS/SO values, providing evidence that the gas-phase C/O ratio commonly exceeds 1 in their outer regions. We present new NOEMA observations of CS $5-4$, SO $7_6-6_5$ and $5_6-4_5$, C$_2$H $N=3-2$, HCN $3-2$, HCO$^+$ $3-2$, and H$^{13}$CO$^+$ $3-2$ in the DR Tau protoplanetary disk at a resolution of $\sim0.4''$ (80 au). Estimates for the disk-averaged CS/SO ratio range from $\sim0.4-0.5$, the lowest value reported thus far for a T Tauri disk. At a projected separation of $\sim180$ au northeast of the star, the SO moment maps exhibit a clump that has no counterpart in the other lines, and the CS/SO value decreases to $<0.2$ at its location. Thermochemical models calculated with DALI indicate that DR Tau's low CS/SO ratio and faint C$_2$H emission can be explained by a gas-phase C/O ratio that is $<1$ at the disk radii traced by NOEMA. Comparisons of DR Tau's SO emission to maps of extended structures traced by $^{13}$CO suggest that late infall may contribute to driving down the gas-phase C/O ratio of its disk. 
\end{abstract}


\section{Introduction \label{sec:intro}}
The C/O ratios of exoplanet atmospheres are commonly compared to protoplanetary disk chemistry models to infer where and how planets formed in their natal disks \citep[e.g.,][]{2011ApJ...743L..16O, 2012ApJ...758...36M, 2021AJ....162..290R, 2022ApJ...934...74M}. Understanding how C/O ratios vary within disks, between disks, and as a function of evolutionary stage is essential for benchmarking planet formation models. 

The most commonly used method to estimate the gas-phase C/O ratio of disks has been to compare millimeter wavelength C$_2$H observations to disk chemistry models in which the elemental C/O ratio is a free parameter \citep[e.g.,][]{2016ApJ...831..101B, 2019AA...631A..69M, 2021ApJS..257....7B}. Millimeter wavelength line emission generally traces disk radii ranging from tens to hundreds of au and disk heights ranging from $z/r\sim0.1-0.4$ \citep[e.g.,][]{2017AA...607A.130D, 2021ApJS..257....4L, 2023AA...669A.126P}. As C/O values rise, production of hydrocarbons such as C$_2$H increases. A C/O value of 1 corresponds to a tipping point at which C$_2$H column densities can change by an order of magnitude or more \citep[e.g.,][]{2018ApJ...865..155C}. Bright C$_2$H emission is commonly detected in Class II disks, suggesting that the gas-phase C/O ratio in their outer regions usually exceeds 1 \citep[e.g.,][]{2016AA...592A.124G, 2019ApJ...876...25B, 2019AA...631A..69M, 2021ApJ...911..150P}. For reference, the C/O ratio of the Sun is 0.55 \citep{2009ARAA..47..481A}. In other words, the disks that have been observed so far typically exhibit carbon-dominated chemistry and super-solar gas-phase C/O ratios. 

Another approach is to compare measurements of the CS/SO column density ratio to disk chemistry models with different input C/O ratios \citep[e.g.,][]{2011AA...535A.104D, 2018AA...617A..28S, 2021ApJS..257...12L, 2023NatAs...7..684K}. CS/SO increases as the elemental C/O ratio increases, again with a significant jump occurring at the critical C/O value of 1. Most attempts to observe SO so far in Class II disks have yielded non-detections, which likewise have been interpreted as evidence that the gas-phase C/O ratio in these systems is $>1$ \citep[e.g.,][]{2011AA...535A.104D, 2013AA...549A..92G, 2018AA...617A..28S, 2021ApJS..257...12L,2021AJ....162...99F}. 

In the simplest picture of planet formation via core accretion, the C/O ratio of a planet's atmosphere reflects the composition of the gas that the planet accreted from its natal protoplanetary disk \citep[e.g.,][]{2011ApJ...743L..16O}. The high gas-phase C/O ratios typically reported for disks are thus somewhat puzzling in light of many wide-separation giant exoplanet atmospheres having C/O ratios less than 1 \citep[e.g.,][]{2023AJ....166...85H}. Explanations invoked to address this apparent discrepancy include planet formation through gravitational instability or accretion of icy planetesimals that increase the atmospheric oxygen levels \citep[e.g.,][]{2021ApJS..257....7B}. Other possibilities are that the C/O ratios of the disk layers traced by millimeter wavelength observations do not necessarily reflect the composition of the midplane or that the bulk of the planet material was accreted when the C/O ratio was lower \citep[e.g.,][]{2023AA...674A.211C}. 

However, SO detections are now being increasingly reported in Class II disks, suggesting that there may be a sub-population of disks that maintains a substantial reservoir of oxygen-dominated gas \citep[e.g.,][]{2016AA...589A..60P, 2018AA...611A..16B, 2021AA...651L...6B, 2023ApJ...943..107H, 2023ApJ...952L..19L}. In nearly all cases, the reported SO detections have come from transition disks hosted by (intermediate-mass) Herbig Ae stars. \citet{2021AA...651L...6B, 2023AA...678A.146B} suggested that such systems exhibit strong SO emission because the highly luminous stellar hosts heat the cavity walls to the point where oxygen-rich ices readily sublimate. 

Among the Class II systems with published SO detections, DR Tau stands out as a T Tauri star that hosts a disk without any indication of a central cavity in images down to a resolution of $\sim0.1''$ ($\sim20$ au) \citep{2019ApJ...882...49L, 2021ApJ...908...46B, 2023ApJ...943..107H}. DR Tau is located 192 pc away in the Taurus star-forming region \citep{1949ApJ...110..424J, 2021AJ....161..147B}. CO, [C I], and scattered light observations have unveiled envelope-like and large-scale spiral structures associated with its disk, indicating that it is likely undergoing late infall \citep{2022AA...660A.126S, 2022AA...658A..63M, 2023ApJ...943..107H}. The detection of SO in the DR Tau disk suggests that oxygen-dominated gas-phase chemistry may persist under a more diverse range of circumstances than implied by the previous detections in Herbig Ae transition disks. However, SO emission by itself does not constrain the C/O ratio, since it is also dependent upon the total sulfur abundance. In order to constrain the C/O ratio of the DR Tau disk, we used the Northern Extended Millimeter Array (NOEMA) to obtain spatially resolved images of CS, SO, and C$_2$H emission and compared these observations with thermochemical models generated using the DALI code \citep{2012AA...541A..91B, 2013AA...559A..46B}. The observations and data reduction are presented in Section~\ref{sec:observations}, and a descriptive analysis of the line observations is provided in Section~\ref{sec:molecularlines}. The modeling procedure and results are provided in Section~\ref{sec:models}. The results are discussed in Section~\ref{sec:discussion} and summarized in Section~\ref{sec:summary}.

\section{Observations and Data Reduction}\label{sec:observations}
Observations of DR Tau were obtained with the NOEMA PolyFiX correlator in dual polarization mode under program W21BE (PI: J. Huang). The correlator was configured to cover frequencies from $244.2-252.3$ GHz and $259.7-267.8$ GHz at a resolution of 2 MHz, while individual lines of interest within these frequency ranges were covered at a higher resolution of 62.5 kHz ($\sim0.07-0.08$ km s$^{-1}$). The primary species of interest were CS, SO, and C$_2$H. The wide bandwidth of the correlator also enabled us to target auxiliary species, including HCN, HCO$^+$, and H$^{13}$CO$^+$. Molecular data for the targeted transitions are given in Table \ref{tab:detections}. DR Tau was observed for 3 hours on source on 19 December 2021 using 10 antennas with baselines ranging from 24 to 400 m, and then for 3.4 hours on source on 26 February 2022 using 11 antennas with baselines ranging from 32 to 920 m. Both sets of observations employed LkH$\alpha$ 101 as the flux calibrator, 3C84 as the bandpass calibrator, and 0507+179 as a phase calibrator. 0446+112 was included as an additional phase calibrator for the second set of observations. 

\begin{deluxetable*}{ccccc}
\tabletypesize{\scriptsize}
\tablecaption {Molecular data \label{tab:detections}}
\tablehead{\colhead{Transition}&\colhead{Rest Frequency} & \colhead{A$_{ij}$} &\colhead{E$_u$}&\colhead{$g_u$} \\& \colhead{(GHz)} &\colhead{(s$^{-1}$)} &\colhead{(K)}}
\startdata
CS $J = 5 - 4$ & 244.9355565 & 2.98$\times10^{-4}$ & 35.3&11\\
SO $J_{N} = 5_{6}-4_{5}$ & 251.82577 & 1.93$\times10^{-4}$ & 50.7& 11\\
H$^{13}$CO$^+$ $J=3-2$ & 260.2553390 & $1.34\times10^{-3}$&25.0&7\\
SO $J_{N} = 7_{6}-6_{5}$ & 261.843721 & 2.28$\times10^{-4}$ & 47.6&15\\
C${_2}$H $N = 3 - 2$, $J= \frac{7}{2} - \frac{5}{2} $, $F = 4 - 3$ & 262.00426 & 5.32$\times10^{-5}$ & 25.1&9\\
C${_2}$H $N = 3 - 2$, $J =\frac{7}{2} - \frac{5}{2} $, $F = 3 - 2$ & 262.006482 & 5.12$\times10^{-5}$ & 25.1&7\\
C${_2}$H $N = 3 - 2$, $J =\frac{5}{2} - \frac{3}{2} $, $F = 3 - 2$ & 262.064986 & 4.89$\times10^{-5}$ & 25.2&7\\
C${_2}$H $N = 3 - 2$, $J =\frac{5}{2} - \frac{3}{2}$, $F = 2 - 1$ & 262.067469 & 4.47$\times10^{-5}$ & 25.2&5\\
HCN $J = 3 - 2$ & 265.8864339 & 8.36$\times10^{-4}$ & 25.5&21\\
HCO$^+$ $J = 3 - 2$ & 267.5576259 & 1.45$\times10^{-3}$ & 25.7&7\\
\enddata 
\tablecomments{Molecular data from the Cologne Database for Molecular Spectroscopy \citep{2001AA...370L..49M,2005JMoSt.742..215M}. In particular, see \url{https://cdms.astro.uni-koeln.de/} for the $g_u$ conventions. The CDMS values are derived from the following references: CS\textemdash\citet{1968JMoSp..28..266W, Bogey1981, 1982JMoSp..95...35B, 1984CaJPh..62.1414W, 1987JMoSp.124..450B, 1995JMoSp.173..146R, 1999ZNatA..54..131A, 2003JMoSp.219..296K, 2003ApJ...588..655G}, SO\textemdash \citet{1964JChPh..41.1413P, CLARK1976332, 1982JMoSp..91...60T, 1992ApJ...399..325L, 1994JMoSp.167..468C, 1996JMoSp.180..197K, 1997JMoSp.182...85B}, H$^{13}$CO$^+$ and HCO$^+$\textemdash\citet{1984JChPh..81.5239D,1985JChPh..82.1750K, 1988JMoSp.127..527H, FT9938902219, 2001ApJ...553.1042G, 2004AA...419..949S, 2007ApJ...662..771L, 2007ApJ...669L.113T}, HCN\textemdash\citet{1984JChPh..80.3989E, 2000JMoSp.202...67M, 2002ZNatA..57..669A, 2003ApJ...585L.163T, 2006SPIE.6580E..01L}, C$_2$H\textemdash\citet{1981ApJ...251L.119S, 1987JChPh..87...73K, 1987AA...186L..14W, 1988JChPh..89.3962K, 1988JMoSp.131...58K, 1993JChPh..98.6690H, 1995JChPh.103.5919H, 1995CPL...244...45W, 1999JChPh.111.1454C, 2004MolPh.102.2167T, 2007JChPh.127k4320K, 2009AA...505.1199P}} 
\end{deluxetable*}

The observations were calibrated with the NOEMA pipeline in \texttt{GILDAS} \citep{2005sf2a.conf..721P, 2013ascl.soft05010G}, with each spectral window (SPW) written out to a $uv$-table. Continuum $uv-$tables were produced by flagging strong line emission in the wide-bandwidth SPWs and averaging the unflagged channels. Three rounds of phase self-calibration were applied to each of the continuum SPWs using solution intervals of 180, 90, and 45 seconds, respectively. Self-calibration improved the continuum S/N by a factor of 7. The resulting self-calibration tables were then applied to the high-resolution SPWs that fell within the same basebands. Continuum subtraction was performed in the $uv-$plane for each SPW by fitting a linear baseline. 

The processed $uv$-tables were then converted to measurement sets for imaging in Common Astronomy Software Applications (CASA) 6.4 \citep{2022PASP..134k4501C}. Line images were produced with channel spacings of 0.2 km s$^{-1}$ using multi-scale CLEAN \citep{2008ISTSP...2..793C} with scales of [$0''$, $0\farcs25$, $0\farcs5$]  and circular masks with a radius of $2''$. The robust value was set to 0.5. To improve the SNR of the weak C$_2$H lines, we employed a $uv$ taper of $0\farcs5$. Primary beam corrections were then applied after CLEANing. Channel maps are provided in Appendix~\ref{sec:chanmaps}.

Moment 0 (integrated intensity) maps were created for HCO$^+$ by integrating between LSRK velocities of 6.6 and 12.4 km s$^{-1}$ and for the other lines by integrating between LSRK velocities of 9.2 and 10.6 km s$^{-1}$. The velocity ranges for HCO$^+$ and CS were chosen based on the channels in which emission was detected above the $3\sigma$ level, where $\sigma$ is the rms measured in nearby line-free channels. The velocity range used to create the moment maps for the weaker lines was set to the values used for CS. We used CS rather than HCO$^+$ to set the velocity range for the other lines because HCO$^+$ has a contribution from non-disk emission (see Section \ref{sec:molecularlines}). Spectra were extracted from the image cubes using circular apertures with radii of $1.5''$, chosen to cover the extent of emission detected above $3\sigma$. Line fluxes were measured from the spectra by integrating through the same velocity range used to make the moment maps. The flux uncertainties were estimated as $\sqrt{N}\Delta v\sigma_\text{spec}$, where $N$ is the number of channels included in the integrated intensity map, $\Delta v$ is the channel spacing, and $\sigma_\text{spec}$ is the standard deviation of a line-free portion of the spectrum. We also created moment maps of the emission redshifted and blueshifted with respect to the systemic velocity of 9.9 km s$^{-1}$ \citep{2021ApJ...908...46B} to compare their kinematics. The synthesized beam, rms, and flux values are listed in Table~\ref{tab:imageproperties}. 

\begin{figure}
\begin{center}
\includegraphics{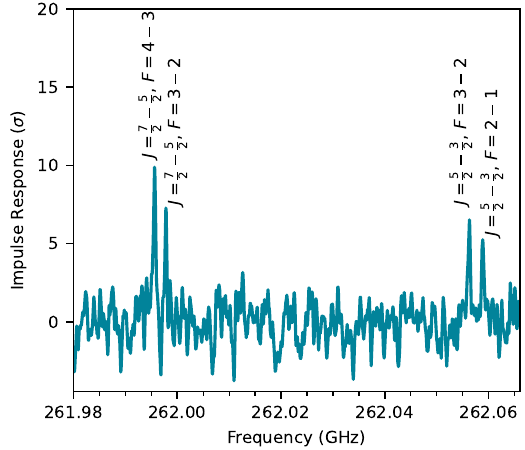}
\end{center}
\caption{Matched filter impulse response spectrum labelled with the four detected C$_2$H $N=3-2$ hyperfine components.  \label{fig:C2Hmatchedfilter}}
\end{figure}

Because the individual hyperfine components of C$_2$H are not clearly detected in the image cubes, we stacked the visibilities (with each component equally weighted) and imaged them with a $uv$ taper of $0\farcs3$. The stacked C$_2$H is detected at $>5\sigma$ in one channel and $>3\sigma$ in an additional four channels (see Appendix~\ref{sec:chanmaps}). To further validate the detection, we used the matched filter method implemented in the \texttt{VISIBLE} Python package \citep{2018AJ....155..182L, 2020ApJ...893..101L}. \texttt{VISIBLE} cross-correlates a set of visibilities with a template that has an emission distribution similar to that expected of the target line in order to determine whether the visibilities contain the signal of interest. The CLEAN model for CS was selected as the matched filter template because this line is detected at high significance and does not have obvious emission originating from non-disk components. The resulting impulse response spectrum is shown in Figure~\ref{fig:C2Hmatchedfilter}, confirming that the four targeted C$_2$H hyperfine components are detected above $5\sigma$.   

Calibrated visibilities and images can be downloaded at \url{https://zenodo.org/records/12600447}.

\begin{deluxetable*}{lcccc}
\tablecaption{Imaging Summary \label{tab:imageproperties}}
\tablehead{
\colhead{Transition}&\colhead{Synthesized beam}&\colhead{Per-channel rms}&\colhead{Moment 0 rms}&\colhead{Flux\tablenotemark{a}}\\
&(arcsec $\times$ arcsec ($^\circ$))&(mJy beam$^{-1}$)&(mJy beam$^{-1}$ km s$^{-1}$)&(mJy km s$^{-1}$)}
\startdata
\multicolumn{5}{c}{Primary line targets}\\
\hline
CS $J=5-4$ & $0\farcs62\times0\farcs32$ ($12\fdg4$)&5.9&4.1&$385\pm10$\\
SO $J_N = 7_6-6_5$ & $0\farcs58\times0\farcs29$ ($12\fdg1$) &6.1&4.0& $231\pm11$\\
SO $J_N = 5_6-4_5$ & $0\farcs61\times0\farcs30$ ($11\fdg8$)&5.4&3.9&$141\pm10$\\ 
C$_2$H\tablenotemark{b} $N=3-2$, $J=\frac{7}{2}-\frac{5}{2}$, $F=4-3$ & $0\farcs69\times0\farcs49$ ($13\fdg7$) &6.5 &4.3& $<30$\\
C$_2$H $N=3-2$, $J=\frac{7}{2}-\frac{5}{2}$, $F=3-2$ & $0\farcs69\times0\farcs49$ ($13\fdg8$) & 6.5& 4.5& $<30$\\
C$_2$H $N=3-2$, $J=\frac{5}{2}-\frac{3}{2}$, $F=3-2$ & $0\farcs69\times0\farcs49$ ($13\fdg8$) & 6.5&4.6 & $<30$\\
C$_2$H $N=3-2$, $J=\frac{5}{2}-\frac{3}{2}$, $F=2-1$ & $0\farcs69\times0\farcs49$ ($13\fdg8$) &6.5 & 4.3 & $<30$\\
C$_2$H  stacked & $0\farcs61\times0\farcs39$ ($12\fdg9$) &2.8 &1.8 & $24\pm5$  \\ 
\hline
\multicolumn{5}{c}{Auxiliary line targets}\\
\hline
HCO$^+$ $J=3-2$ & $0\farcs56\times0\farcs28$ ($12\fdg4$)&8.2&14& $2220\pm40$\\
H$^{13}$CO$^+$ $J=3-2$ &  $0\farcs58\times0\farcs29$ ($11\fdg9$)&6.0&4.0&$122\pm12$\\
HCN $J=3-2$ & $0\farcs57\times0\farcs29$ ($12\fdg3$)&6.6&4.4&$143\pm11$\\
\enddata
\tablenotetext{a}{The $1\sigma$ error bars do not include the $\sim10\%$ systematic flux uncertainty. Upper limits are $3\sigma$. }
\tablenotetext{b}{While emission from the C$_2$H $N=3-2$ hyperfine components is often blended in disks, the nearly face-on inclination of DR Tau allows the components to be separated.}
\end{deluxetable*}

\section{Overview of line observations \label{sec:molecularlines}}
\subsection{Emission Morphology}
\begin{figure*}
\begin{center}
\includegraphics[width = 0.95\textwidth]{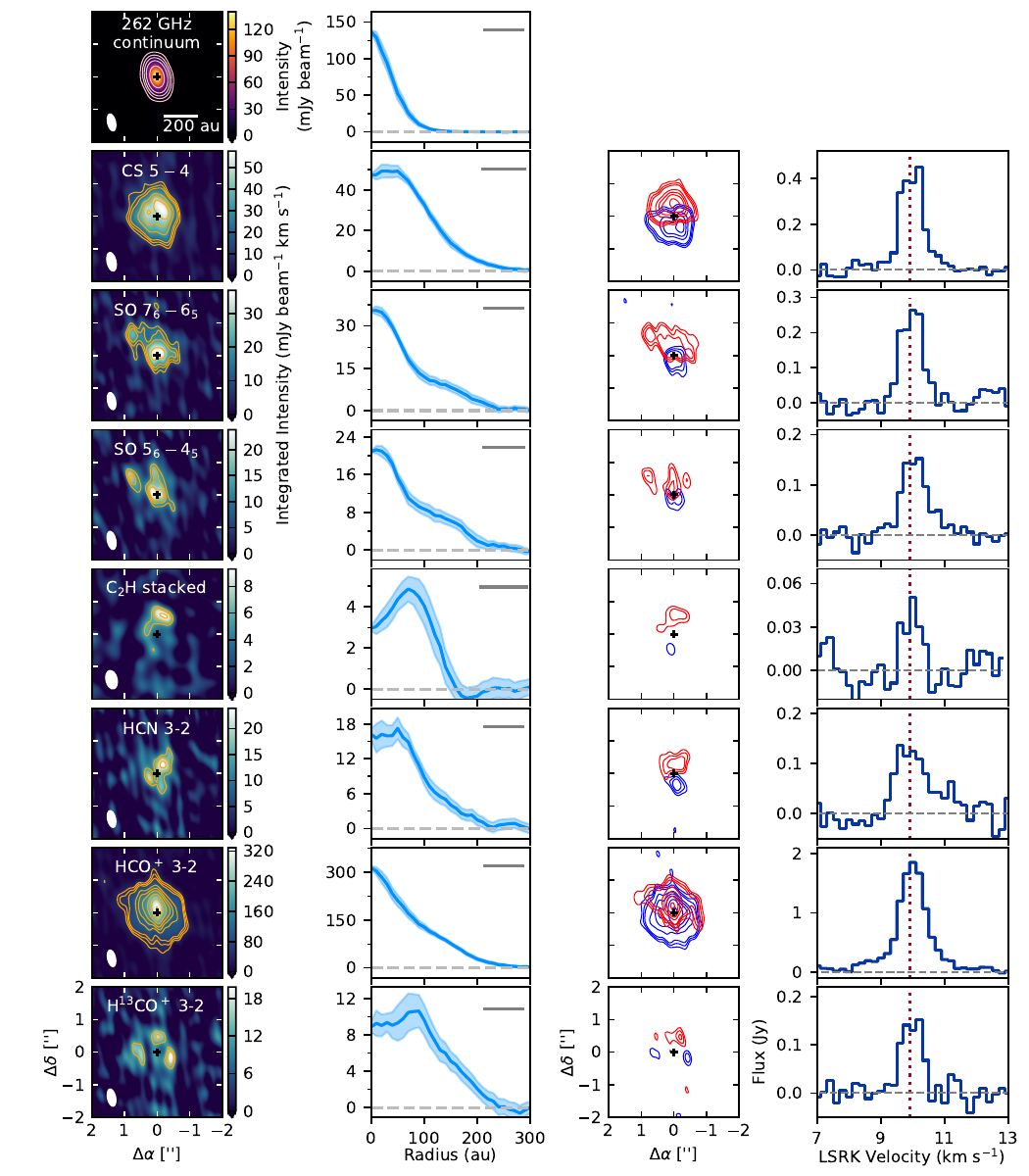}
\end{center}
\caption{First column: 262 GHz continuum and integrated intensity maps. The continuum contours are drawn at [10, 25, 50, 100, 250, 500]$\sigma$, where $\sigma=0.25$ mJy beam$^{-1}$. The integrated intensity map contours are drawn at [3, 4, 5, 8, 10, 12, 15, 20, 25]$\sigma$, where $\sigma$ is listed in Table~\ref{tab:imageproperties} for each line. The synthesized beam is drawn in the lower left corner of each panel. The black crosses mark the location of the continuum peak. Second column: Deprojected, azimuthally averaged radial profiles. The blue shading shows the $1\sigma$ error, where $\sigma$ is calculated by dividing the scatter in each radial bin by \edit1{$\max\left(1, \sqrt{N}\right)$}, where $N$ is the number of beams spanned by the corresponding annulus. The gray bars show the geometric mean of the major and minor axes of the synthesized beam. Third column: Moment maps of the emission blueshifted and redshifted with respect to the systemic velocity (9.9 km s$^{-1}$). Contours are drawn at [3, 4, 5, 8, 10, 12, 15, 20]$\sigma/\sqrt{2}$, where $\sigma$ is the noise level of the full moment maps (which are integrated over $2\times$ as many channels as the redshifted and blueshifted maps). Fourth column: Corresponding spectra. The dotted red lines denote the systemic velocity. \label{fig:observationoverview}}
\end{figure*}

The integrated intensity maps, maps of redshifted and blueshifted emission, radial intensity profiles, and spectra are presented in Figure~\ref{fig:observationoverview}. 
Most molecules exhibit redshifted emission to the north of the star and blueshifted emission to the south, consistent with expectations for a rotating disk and with the kinematics of previous observations of C$^{18}$O, H$_2$CO, and SO \citep{2023ApJ...943..107H}. The exception is HCO$^+$, for which the blueshifted emission encircles the star, and the redshifted emission exhibits protrusions south of the star. This suggests that some of the HCO$^+$ emission is coming from the envelope and/or outflow material previously detected in $^{12}$CO, $^{13}$CO, and [C I] \citep{2022AA...660A.126S, 2023ApJ...943..107H}.

\begin{figure*}
\begin{center}
\includegraphics{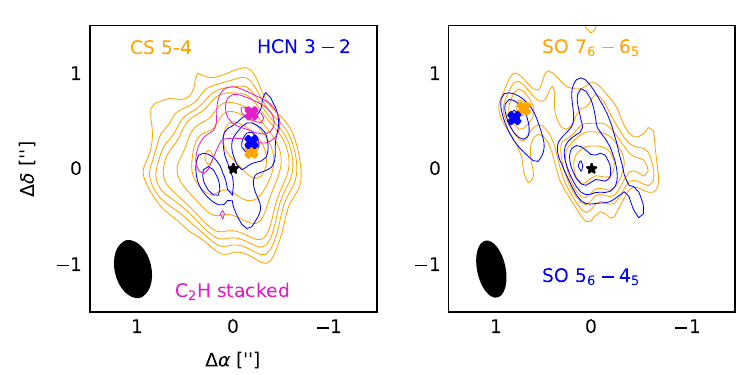}
\end{center}
\caption{Left: Overlaid contour maps of CS (orange), HCN (blue), and C$_2$H (magenta) showing the positions of the emission asymmetries. Contours are drawn at [3, 4, 5, 6, 8, 10, 12, 14]$\sigma$. The black star marks the location of DR Tau. The orange, blue, and magenta crosses mark the emission peaks of CS, HCN, and C$_2$H, respectively. Right: Similar to left panel, except for SO $7_6-6_5$ (orange) and SO $5_6-4_5$ (blue). The crosses mark the emission peaks of the northeast clump for each transition. \label{fig:overlaidmaps}}
\end{figure*}

Several molecules show evidence of asymmetry in their integrated intensity maps (Figure~\ref{fig:overlaidmaps}). CS, HCN, and the stacked C$_2$H emission all peak to the northwest of the star, although it is less certain whether the apparent HCN and C$_2$H asymmetries are real because their peak intensities are only at the $\sim5\sigma$ level. CS peaks at a projected separation of $\sim50$ au (since DR Tau is nearly face-on, the deprojected separation is essentially the same), HCN peaks at a separation of $\sim65$ au, and C$_2$H peaks at a separation of $\sim120$ au. The difference in their peak positions, however, is smaller than the synthesized beam, so observations at higher resolution and sensitivity will be necessary to confirm the apparent offsets. Meanwhile, the moment maps for SO $7_6-6_5$ and $5_6-4_5$ each show two emission components: the first consists of bright emission centered on the star, and the other is a clump-like structure at a separation of $\sim180$ au from DR Tau and a P.A. of $\sim50^\circ$. \citet{2023ApJ...943..107H} previously noted an asymmetry to the northeast in SO $6_5-5_4$ emission, but did not resolve the separate components. The clump does not have a clear counterpart in any other line observed toward DR Tau, suggesting that it traces a localized chemical change rather than a gas overdensity. It is located outside both the millimeter continuum and the structures observed with SPHERE in polarized scattered light \citep{2022AA...658A..63M}. The SO $5_6-4_5$ clump is shifted slightly southeast of the SO $7_6-6_5$ clump. Given that the two transitions have similar upper-state energy levels (50.7 K for SO $5_6-4_5$ versus 47.6 K for SO $7_6-6_5$), it seems unlikely that the emission morphology difference is due to excitation differences. The apparent separation in the northeast clump positions is smaller than that of the synthesized beam, and their signal-to-noise ratios are only $\sim5-6\sigma$, so higher-resolution and deeper observations will be needed to ascertain whether the clumps are genuinely offset in the two transitions. 

\begin{figure*}
\begin{center}
\includegraphics{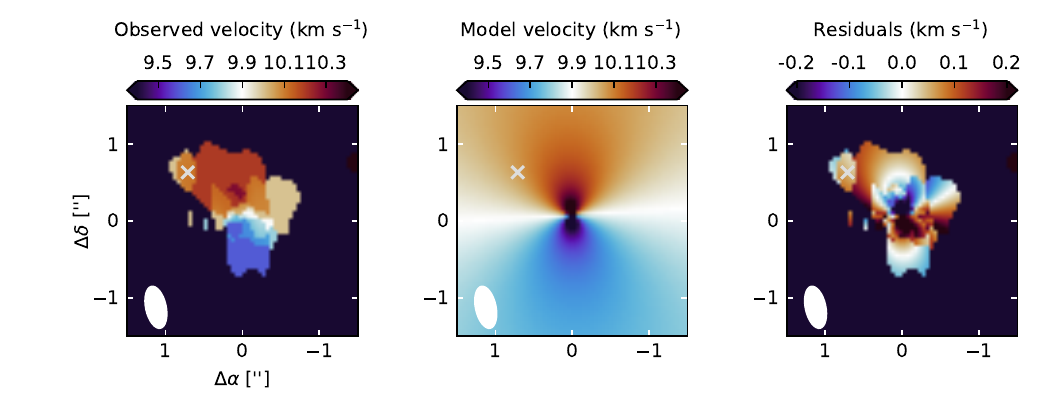}
\end{center}
\caption{Left: SO $7_6-6_5$ intensity-weighted velocity map. The gray cross marks the position of the northeast clump. Center: Model Keplerian velocity map. Right: Velocity residual map. \label{fig:SOmom1}}
\end{figure*}

We attempted to fit the velocity map for SO $7_6-6_5$ (the SO transition detected at higher SNR) with a Keplerian model using \texttt{eddy} \citep{2019JOSS....4.1220T} to determine whether the clump was associated with any non-Keplerian motion, but the data quality was not sufficient to obtain a reliable fit. Instead, we generated a model Keplerian velocity map with \texttt{eddy} using parameters measured from previous observations of DR Tau: $v_\text{sys}=9.9$ km s$^{-1}$, $M_\star=1.18$ $M_\odot$ \citep{2021ApJ...908...46B}, $i=5.4^\circ$, and P.A. $=3.4^\circ$ \citep{2019ApJ...882...49L}. The model Keplerian map was then subtracted from the observed SO map (Figure~\ref{fig:SOmom1}). The line-of-sight velocity clump differs from the Keplerian model by $\sim$ 50 m s$^{-1}$, which is smaller than the spectral resolution. Thus, the clump does not appear to exhibit significant deviations from Keplerian motion, but observations at greater sensitivity and resolution, both spectral and spatial, are needed to characterize the kinematics in a rigorous fashion. The literature values of $M_\star$, $i$, and P.A. have large uncertainties due to the low inclination and compactness of the disk. In addition, because the disk is nearly face-on, we are primarily sensitive to deviations from Keplerian motion in the vertical direction rather than the radial or azimuthal directions. Thus, a large absolute deviation from Keplerian motion could still yield only a small line-of-sight deviation.

The radial profiles of H$^{13}$CO$^+$ and C$_2$H suggest that their emission is arranged in ring-like structures, with the radial intensity profile of the former peaking at $\sim90$ au and the latter at $\sim70$ au. The HCO$^+$ emission is centrally peaked rather than ring-like because HCO$^+$ is more optically thick than H$^{13}$CO$^+$.

\subsection{The CS/SO ratio}
Under the assumption that the molecular emission is in LTE and optically thin, the column density $N$ can be estimated by 

\begin{equation}
N = \frac{4\pi S_\nu\Delta v}{A_{ul}\Omega h c} \frac{Q(T)}{g_u}e^{E_u/T},
\end{equation}
where $S_\nu\Delta v$ is the velocity-integrated flux, $A_{ul}$ is the Einstein-A coefficient of the transition, $\Omega$ is the solid angle subtended by the area over which the flux is measured, $T$ is the gas temperature, $Q$ is the partition function, $g_u$ is the upper state degeneracy, and $E_u$ is the upper state energy level \citep[see, e.g.,][]{1999ApJ...517..209G, 2008AA...488..959B}. The values for $A_{ul}$, $g_u$, and $E_u$ are taken from the Cologne Database for Molecular Spectroscopy \citep{2001AA...370L..49M,2005JMoSt.742..215M}, and $Q(T)$ is computed by interpolating the values from the database. 

We first estimated the source-averaged CS/SO ratio using the CS $5-4$ and SO $7_6-6_5$ flux measurements in Table~\ref{tab:imageproperties}. SO $7_6-6_5$ was selected rather than $5_6-4_5$ for these estimates because the former is detected at higher SNR. Because the upper state energies of the two SO transitions are very close to one another (47.6 K and 50.7 K, respectively), they do not meaningfully constrain the excitation temperature. We therefore calculated column densities for assumed temperatures of 30 K and 80 K, chosen based on gas temperatures estimated from CO brightness temperature measurements in other disks hosted by T Tauri stars \citep[e.g.,][]{2021ApJS..257....4L}. For an assumed $T=30$ K, we estimate that $N_\text{CS} = 3.8\pm0.4 \times 10^{12}$ cm$^{-2}$, $N_\text{SO} = 8.8\pm1.0\times 10^{12}$ cm$^{-2}$, and CS/SO $=0.43\pm0.02$.  For an assumed $T=80$ K, we estimate that $N_\text{CS} = 4.8\pm0.5 \times 10^{12}$ cm$^{-2}$, $N_\text{SO} = 9.9\pm1.1\times 10^{12}$ cm$^{-2}$, and CS/SO $=0.48\pm0.03$. The $1\sigma$ uncertainties for the column densities account for the $\sim10\%$ systematic flux calibration uncertainty. However, since CS and SO were observed simultaneously, the flux calibration uncertainties cancel out when taking the CS/SO ratio.   

\begin{figure*}
\begin{center}
\includegraphics{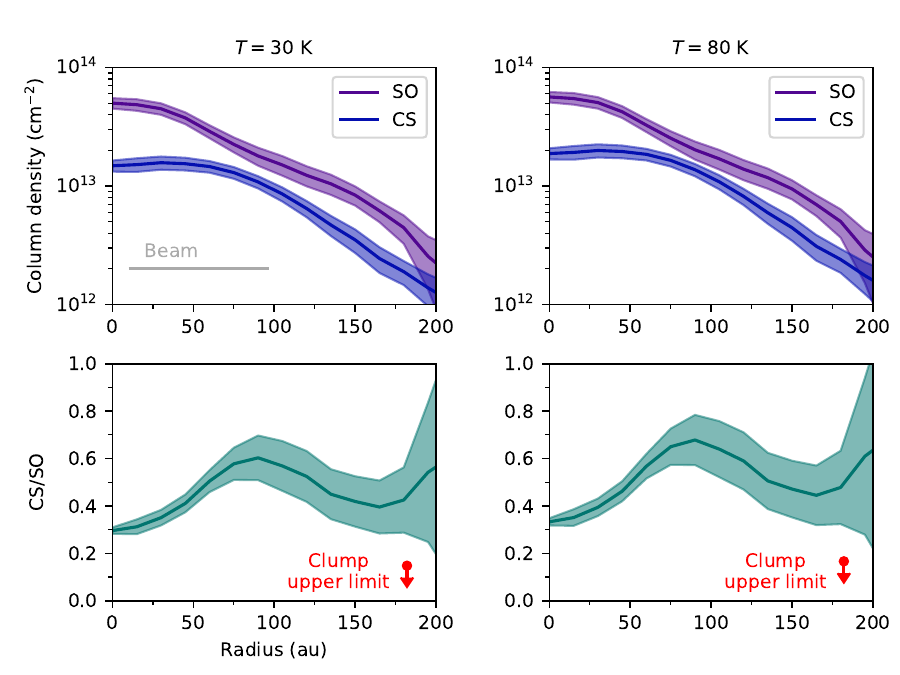}
\end{center}
\caption{Top: Azimuthally averaged CS and SO radial column density profiles (excluding the northeast SO clump) for assumed gas temperatures of 30 and 80 K, respectively. The horizontal gray bar in the upper left panel shows the geometric mean of the major and minor axes of the synthesized beam. The shaded ribbon shows the $1\sigma$ uncertainty. Bottom: Azimuthally averaged radial CS/SO profiles (excluding the northeast SO clump) for assumed gas temperatures of 30 and 80 K, respectively. The red arrows denote the CS/SO upper limit measured at the peak of the SO clump. \label{fig:CStoSOradprofiles}}
\end{figure*}

To calculate the CS/SO ratio as a function of disk radius in DR Tau, we re-imaged CS with a robust value of 0 so that its beam size more closely matched that of SO $7_6-6_5$. We then smoothed both image cubes in CASA with \texttt{imsmooth} to a common beam size of $0.59''\times0.29''$~($12^\circ$), created integrated intensity maps, and calculated new azimuthally averaged radial intensity profiles. The SO radial intensity profile was computed only with pixels located at position angles between 90 and 360$^\circ$ in order to exclude the northeast clump. The results are plotted in Figure~\ref{fig:CStoSOradprofiles}. At $T=30$ K, the CS/SO values range from a minimum of $0.3$ at the disk center to a maximum of $0.6$ at a radius of $90$ au. At $T=80$ K, the estimated CS/SO values are slightly higher, ranging from $0.33$ to $0.68$. The CS/SO ratio appears to decrease slightly beyond 90 au, although the uncertainty is large at these radii and deeper observations will be required to confirm this behavior. The apparent peak of the CS/SO radial profile appears to occur relatively close to the peak of the C$_2$H radial profile (at $\sim70$ au), which may be a reflection of both C$_2$H and CS/SO being sensitive to spatial variations in the gas-phase C/O ratio. The current data, though, are not well-resolved spatially, so higher resolution observations will be needed to study how well the C$_2$H emission morphology correlates with the CS/SO ratio. 

We then estimated the CS/SO ratio at the peak of the northeast SO clump. At $T=30$ K, $N_\text{SO}=3.5\pm0.7 \times10^{13}$, the $3\sigma$ upper limit for $N_\text{CS}$ is $5\times10^{12}$ cm$^{-2}$, and the $3\sigma$ upper limit for CS/SO is $<0.15$. At $T=80$ K, $N_\text{SO}=3.9\pm0.7 \times10^{13}$, the $3\sigma$ upper limit for $N_\text{CS}$ is $6\times10^{12}$ cm$^{-2}$, and the $3\sigma$ upper limit for CS/SO is $<0.17$. (Note that the CS column density upper limit at the clump is higher than the value at the corresponding radius in the radial column density profile because the uncertainties of the latter are decreased through azimuthal averaging). The upper limits are plotted in Figure~\ref{fig:CStoSOradprofiles}. The CS/SO upper limit at the clump is about $2\times$ lower than the minimum CS/SO value measured in the radial CS/SO profile. 

The presence of the clump in SO emission and its absence in CS emission suggest that the CS/SO ratio varies azimuthally, a behavior that has only previously been observed in the HD 100546 disk \citep{2023NatAs...7..684K}. Due to the modest SNR of our observations, we refrain from making a two-dimensional map of the CS/SO ratio. Nevertheless, this source is an excellent candidate for deeper observations to study both azimuthal and radial variations in the CS/SO ratio. 

\section{Thermochemical modeling\label{sec:models}}
\subsection{Modeling DR Tau's disk physical structure}

Employing a procedure adapted from that of \citet{2022AA...660A.126S}, we used the DALI thermochemical code \citep{2012AA...541A..91B, 2013AA...559A..46B} to model DR Tau's SED, spatially resolved millimeter continuum, and C$^{18}$O $J=2-1$ images in order to constrain the disk density and temperature. The modeling procedure is described in more detail in Appendix \ref{sec:modeldescription}.

\subsubsection{Observational constraints}
 The dereddened SED is taken from \citet{2022AA...660A.126S}. Because DR Tau's continuum is not resolved in our NOEMA observations, we retrieved higher resolution 1.3 millimeter continuum observations originally published in \citet{2019ApJ...882...49L} from the ALMA archive. Appendix~\ref{sec:DRTaucontinuum} describes the re-reduction. The NOEMA C$^{18}$O observations were originally published in \citet{2023ApJ...943..107H, huang_2023_7370498}. While \citet{2023ApJ...943..107H} also observed $^{12}$CO and $^{13}$CO 2-1, we elected not to use them to constrain the models because it is not straightforward to distinguish the disk emission from the complex large-scale emission. Our model parametrization neglects these extended structures because they are not detected in C$^{18}$O emission and therefore their mass appears to be small compared to that of the disk. We comment on possible consequences of this choice in Section~\ref{sec:discussion}. Although \citet{2022AA...660A.126S} presented higher resolution ALMA CO isotopologue observations, their maximum recoverable scales are $<2''$, smaller than the scale of molecular emission observed by \citet{2023ApJ...943..107H}. Given that their observations appear to be affected by spatial filtering, we did not include them in our analysis either.

\subsubsection{Physical structure modeling results}
\begin{figure*}
\begin{center}
\includegraphics{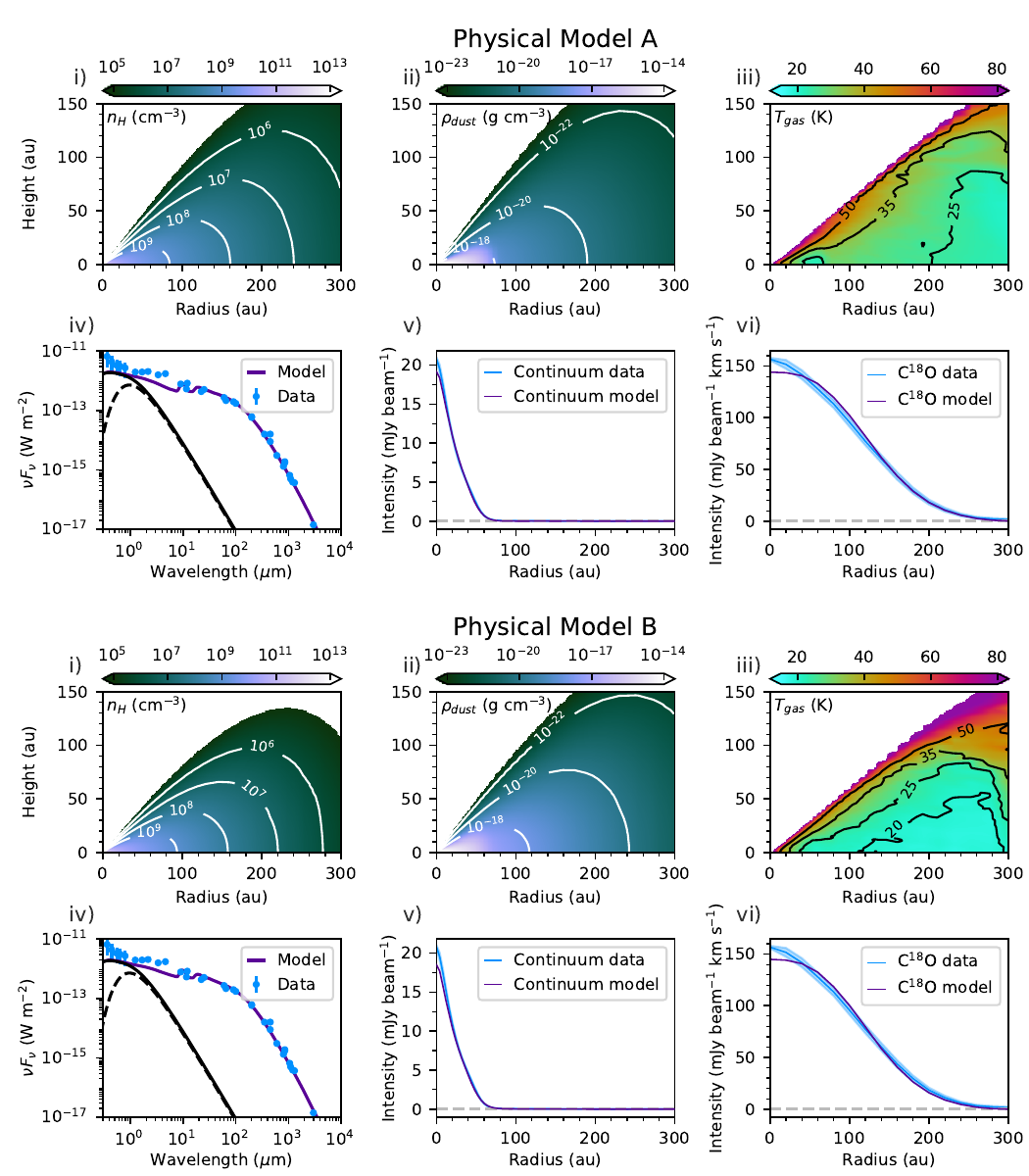}
\end{center}
\caption{Overview of the disk structures corresponding to physical models A (top) and B (bottom). i) Hydrogen nuclei number density. ii) Dust density. iii) Gas temperature. iv) Comparison of DR Tau's dereddened SED to DALI model. The dashed black curve represents the blackbody corresponding to DR Tau's $T_\text{eff}$, while the solid black curve corresponds to the model stellar spectrum that includes a UV excess. v) Comparison of the observed ALMA 1.3 mm continuum radial intensity profile to the DALI model. The light blue shaded region corresponds to the $1\sigma$ uncertainty. vi) Comparison of the observed C$^{18}$O $2-1$ radial intensity profile to the DALI model. \label{fig:physicalstructure}}
\end{figure*}

The final parameter values for models A and B are listed in Table~\ref{tab:diskstructure}. The model density and temperature structures and the corresponding SED, 1.3 mm continuum radial intensity profiles, and C$^{18}$O radial intensity profiles are presented in Figure~\ref{fig:physicalstructure}. The most notable difference between the derived properties for Model A and B is that the latter is colder, since a greater fraction of its mass is in small dust grains, and thus the disk is more opaque to stellar radiation. With a more flexible parametrization, the most notable differences between our models and the DR Tau model presented in \citet{2022AA...660A.126S} is that our value of $r_{c,\text{gas}}$ is about $1.5-2\times$ larger, and we use a steeper value for $q_\text{lg}$. \citet{2022AA...660A.126S} found that their model CO isotopologue emission was too radially compact compared to the observations, so a larger $r_{c,\text{gas}}$ provides a better match to the observations. We find that a high value of $q_\text{lg}$ is necessary to reproduce the rapid fall-off in the continuum radial intensity profile. \citet{2019ApJ...882...49L} similarly concluded that a steep exponential taper was required to reproduce the DR Tau continuum, although they modelled the intensity profile directly rather than performing radiative transfer modelling. Our final models also have an input carbon abundance (carried by CO) that is roughly half that of the model from \citet{2022AA...660A.126S}, which is due in large part to the model gas distribution in \citet{2022AA...660A.126S} being more radially compact and thus requiring a higher CO abundance in order to match the observed spectra. 

\begin{deluxetable}{lcc}
\tablecaption{Disk structure model parameters \label{tab:diskstructure}}
\tablehead{
\colhead{}&\colhead{Physical Model A}&\colhead{Physical Model B}}
\startdata
$r_{subl}$ (au) &0.074 & 0.074  \\
$r_{c,\,gas}$ (au) & 100 & 80 \\
$\gamma_{gas}$ & 0.5 & 0.5\\
$q_{gas}$ & 1.5& 1.5\\
$r_{c,\,lg}$ (au) & 45 & 45\\
$\gamma_{lg}$ & 0.5& 0.5\\
$q_{lg}$ & 4 & 4\\
$M_\text{gas}$ ($M_\odot$) & 0.015 & 0.018 \\
$H_{100}$ (au) & 22 & 19\\
$\psi$ & 1.1 & 1.07\\
$\chi$ & 0.5 & 0.7\\
$f_{lg}$ & 0.99 & 0.90\\
$f_\text{CO}$ & 0.08 & 0.09
\enddata
\end{deluxetable}



\subsection{CS, SO, and C$_2$H modeling}
\subsubsection{Chemical Network}
Since the networks used for modeling the physical structure and C$^{18}$O emission of DR Tau do not include important sulfur and hydrocarbon reactions, we used a modified version of the network from \citet{2023NatAs...7..684K} to model the CS, SO, and C$_2$H observations. In brief, the network from \citet{2023NatAs...7..684K} augments the thermochemistry network from \citet{2013AA...559A..46B} with additional sulfur and hydrocarbon gas-phase reactions from UMIST06 \citep{2007AA...466.1197W}, freezeout, thermal desorption, photodissociation, and photoionization of the additional sulfur-bearing species, and hydrogenation of S to HS and HS to H$_2$S on grains. The largest hydrocarbon in the network is C$_2$H$_3^+$, which serves as a hydrocarbon sink \citep[e.g.,][]{2023AA...673A...7L}. Overall, the network consists of 131 species and 1721 reactions. For this work, the binding energies were updated to the recommended values from \citet{2017ApJ...844...71P}, with the exception of CO, for which we set the binding energy to 855 K to be consistent with the \citet{2016AA...594A..85M} network used to model C$^{18}$O emission. The rate coefficient for the reaction C$_2$H + O $\rightarrow$ CO + CH  was updated to 10$^{-10}$ cm$^3$ s$^{-1}$ from $1.7\times10^{-11}$ cm$^3$ s$^{-1}$ in accordance with UMIST12 \citep{2013AA...550A..36M}, since this was noted as one of the most significant changes from UMIST06. The UMIST12 value is also consistent with the recommended value from the KIDA database \citep{2012ApJS..199...21W}. For the reaction O + HS $\rightarrow$ SO + H, the reaction rate coefficients and temperature range were updated based on the values recommended by \citet{2017MNRAS.469..435V} and the KIDA database: $\alpha = 1.6\times10^{-10}$ cm$^{3}$ s$^{-1}$, $\beta = 0.5$, and $\gamma=0$ K. The key difference is that in UMIST06, the reaction is valid only above 298 K, whereas the reaction is barrierless and valid for temperatures between 10 and 280 K in KIDA. 

The main reservoir of sulfur in disks is not yet clear. In the dense ISM, hypothesized major reservoirs include atomic sulfur, H$_2$S, and organo-sulfur molecules \citep[e.g.,][]{2013ApJ...779..141A, 2017MNRAS.469..435V, 2019AA...624A.108L}. From an analysis of stellar spectra, \citet{2019ApJ...885..114K} inferred that a large fraction of sulfur in disks is in refractory form, most likely FeS. To account for the fact that only some of the sulfur can participate in reactions, chemical models of disks typically allow S/H to be a free parameter \citep{2018AA...617A..28S, 2021ApJS..257...12L, 2024MNRAS.528..388K}. For simplicity, volatile sulfur in our models begins in atomic form. The impact of this assumption is examined later in this section.

\subsubsection{Modeling Procedure} 
For both sets of physical structures A and B, we ran chemistry-only models for 1 Myr each using input elemental C/O ratios of 0.47, 0.7, 0.9, and 1.0. We assumed that all of the C and O are initially in gas-phase CO and H$_2$O ice. The CO initial abundance was fixed to that derived from physical structure modeling ($f_\text{CO}\times1.35\times10^{-4}$ per H), while the H$_2$O ice abundance was scaled to achieve the desired input C/O ratio. The cosmic ray ionization rate was fixed to the value used for the physical structure modeling, 10$^{-18}$ s$^{-1}$, since previous chemical models have shown that the abundances of CS, SO, and C$_2$H are not highly sensitive to the choice of value \citep{2018ApJ...865..155C, 2018AA...617A..28S}. \citet{2018ApJ...865..155C} tested values ranging from $\sim10^{-20}-10^{-18}$ for C$_2$H, while \citet{2018AA...617A..28S} tested values ranging from $\sim10^{-18}-10^{-16}$ for CS and SO.  

Previous comparisons between observations and models of sulfur-bearing species in disks suggest that the total volatile sulfur abundance varies radially \citep[e.g.,][]{2019ApJ...876...72L, 2023NatAs...7..684K}. Whereas CS and SO abundances vary in opposite directions as the C/O value changes, they vary in the same direction with changes in the overall sulfur abundance \citep[e.g.,][]{2018AA...617A..28S}. For each combination of physical structure and input C/O ratio, we estimated the sulfur abundance required to match the CS observations using an approach similar to the iterative procedure described by \citet{2016ApJ...816...25P} for deriving the dust surface density profile of a structured disk. We first ran a model with a guess for the sulfur abundance, raytraced the CS image cube, subtracted the continuum, generated and CLEANed synthetic visibilities, and then extracted the radial intensity profile. Besides S, CO, and H$_2$O, all other initial abundances were fixed to the values listed in Table~\ref{tab:initabundances}. We then performed new model runs after adjusting the input sulfur abundance profile based on the ratio of the observed and model CS radial intensity profiles, which were binned at intervals of 10 au. Since the radial grid points of the DALI models do not coincide with the grid points of the radial intensity profiles, the sulfur abundance of each cell was calculated by linearly interpolating the scaling factors calculated from the ratio of the radial profiles. We were typically able to achieve a good match between the observed and model CS profiles beyond $\sim100$ au within several iterations, but the inner disk always required manual adjustment because it is unresolved and the optical depths are higher. We stopped this procedure after the observed and model CS radial intensity profiles matched within 10\% at radii within 200 au. This then allows us to examine whether SO is over- or under-predicted relative to CS for a given C/O ratio. One could in principle switch the roles of CS and SO in this procedure, but we selected CS as the common reference because of its higher SNR. In general, though, the sulfur abundance profile derived from matching SO will not necessarily be consistent with that derived from matching CS, particularly if the C/O ratio is incorrect. Thus, the derived S abundance is only meaningful when both the CS and SO models are consistent with observations. 

\subsubsection{Chemical modeling results}

Once the input radial sulfur abundance profile was finalized for each combination of physical structure and input C/O ratio, we raytraced the CS $5-4$, SO $7_6-6_5$, and four C$_2$H $N=3-2$ hyperfine components. We again used \texttt{vis\_sample} to generate synthetic visibilities, stacking the C$_2$H visibilities in the same manner as the observations. The CS, SO, and stacked C$_2$H models were CLEANed, and their radial intensity profiles were subsequently extracted from the model integrated intensity maps. For the SO radial profiles, azimuthal angles at P.A.s between 0 and $90^\circ$ were excluded, since the models did not include the northeast clump. For each model, we also computed the local gas-phase C/O ratio by summing over all the carbon and oxygen contained in gas-phase species. In the intermediate and upper layers of the disk, the gas-phase C/O ratio matches the input C/O ratio because the primary C and O carriers, H$_2$O and CO, are in the gas-phase. At lower heights, the gas-phase C/O ratio tends toward 1 because H$_2$O is locked up in ice and CO is the primary carrier of both gas-phase carbon and oxygen. 

\begin{figure*}
\centering
\includegraphics{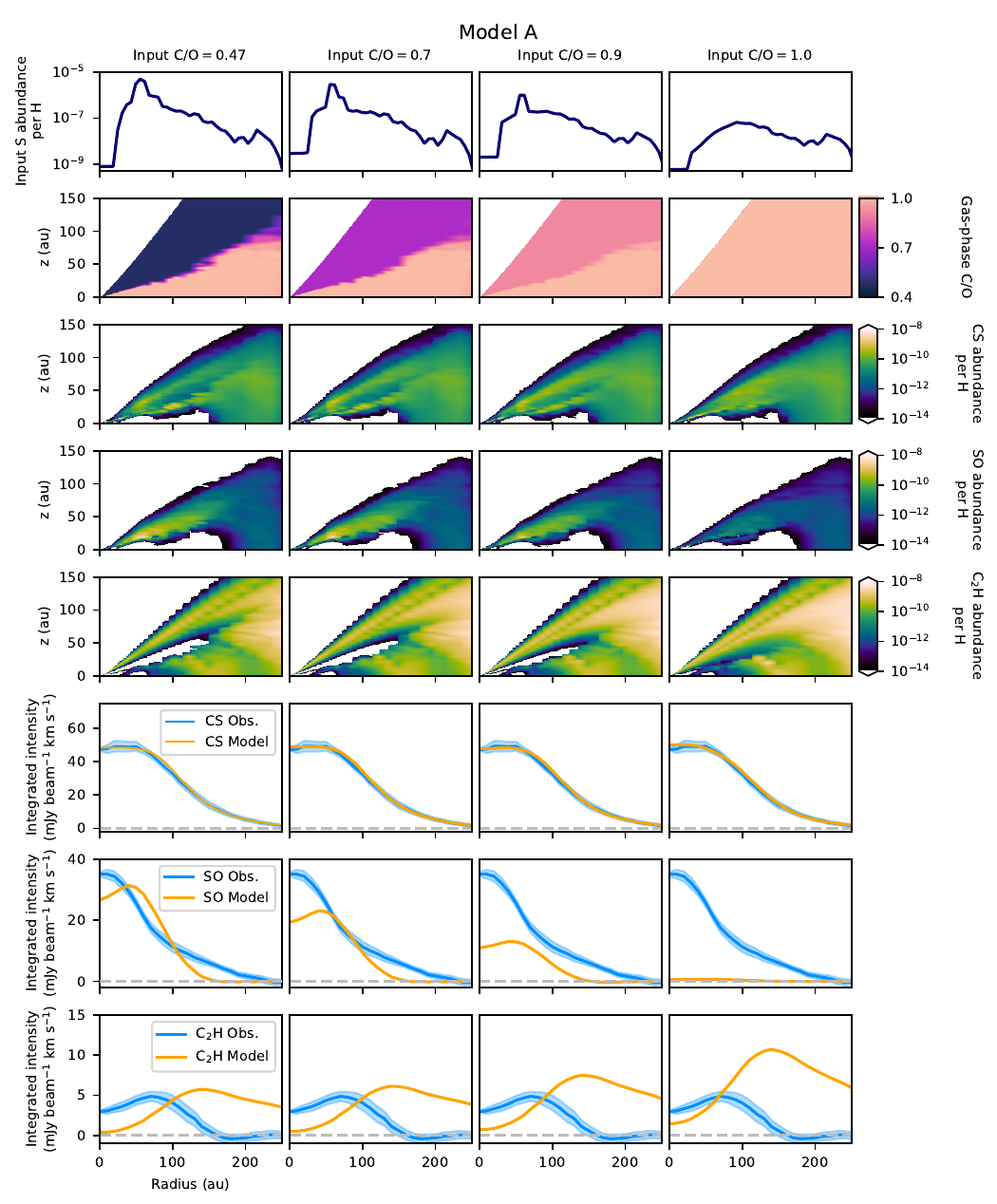}
\caption{a) Chemical model results for physical model A. First row: Input S abundance (relative to hydrogen nuclei) required to match the CS radial intensity profile for each input C/O ratio. Second row: 2D gas-phase elemental C/O ratios at the end of the model run. Third to fifth rows: 2D CS, SO, and C$_2$H model abundances. Sixth to eighth rows: Comparison of observed and model radial integrated intensity profiles for CS $5-4$, SO $7_6-6_5$, and stacked C$_2$H. \label{fig:chemmodels}}
\end{figure*}

\begin{figure*}
\begin{center}
\ContinuedFloat
\includegraphics{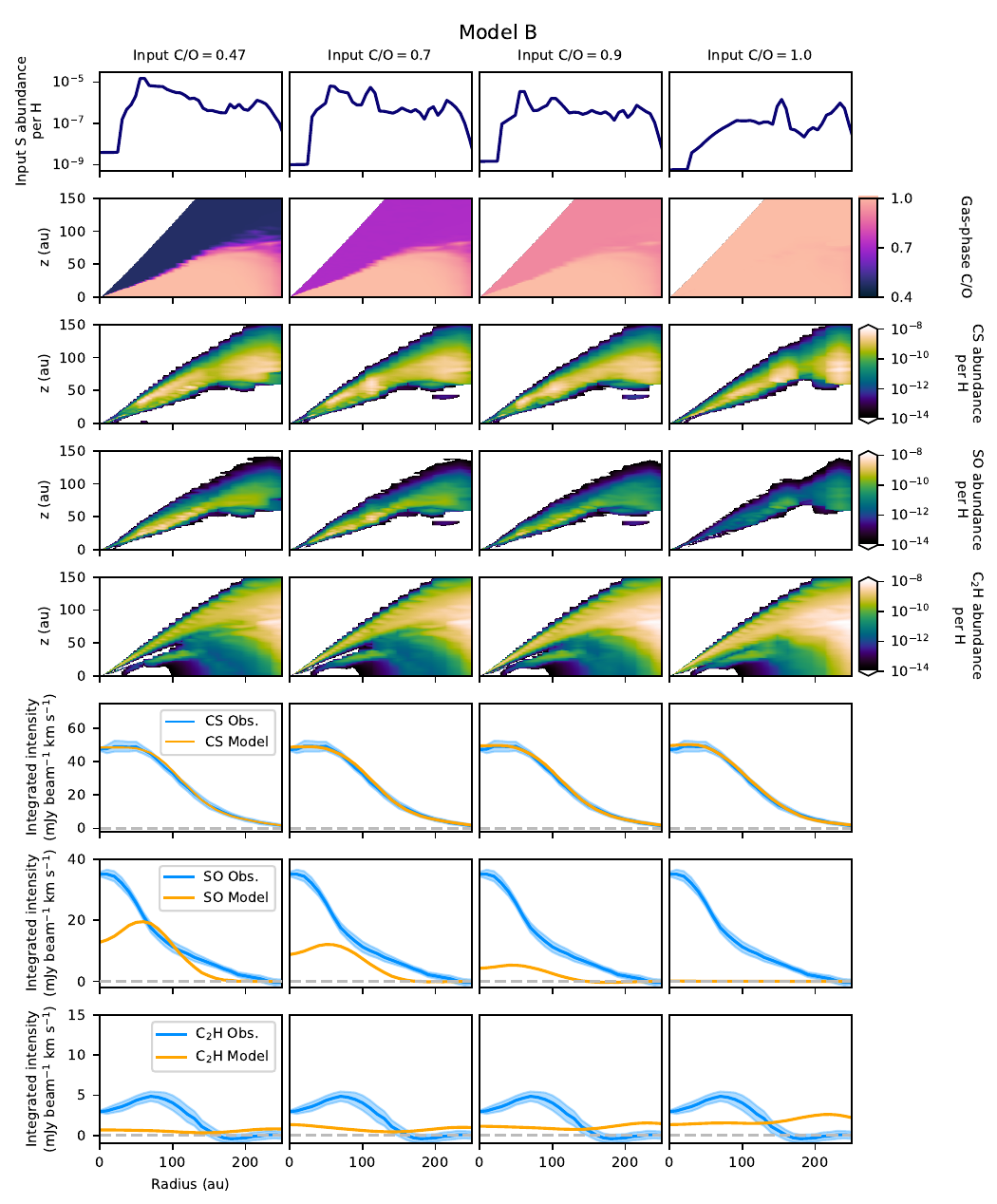}
\caption{b) Similar to a), except using physical model B.}
\end{center}
\end{figure*}

\begin{figure*}
\centering
\includegraphics{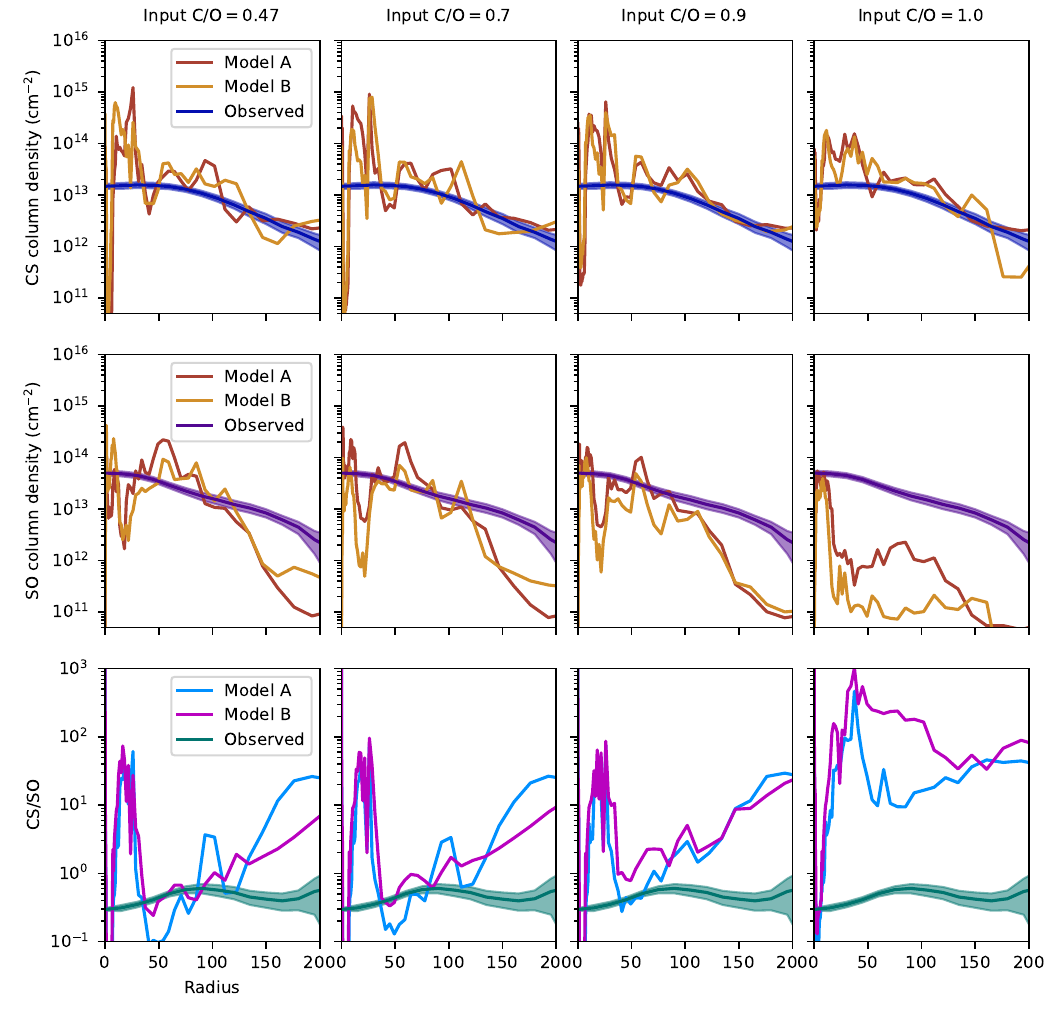}
\caption{Comparisons of the observed and model CS and SO column densities and CS/SO ratios for different input C/O ratios. The observed quantities correspond to the values from Figure~\ref{fig:CStoSOradprofiles} for an assumed excitation temperature of 30 K. \label{fig:coldensitycomparisons}}
\end{figure*}

The chemical model results and comparisons with the observed radial intensity profiles are shown in Figure~\ref{fig:chemmodels}. Comparisons between the observed and model CS and SO column densities and CS/SO ratio are shown in Figure~\ref{fig:coldensitycomparisons}. Outside a radius of 50 au, the CS column densities obtained from matching the DALI models to the radial intensity profile correspond well to the column densities calculated directly from the radial intensity profile. However, the DALI column densities are one to two orders of magnitude higher interior to 50 au. The discrepancy in the inner disk is due to a combination of the higher optical depths and beam smearing. Thus, the CS/SO ratio estimated in the inner 50 au may not be reliable. In general, the input S abundance required to match the CS emission decreases as the input C/O increases because a larger fraction of the sulfur budget gets incorporated into CS. In addition, as the input C/O increases, the SO abundance decreases, while the C$_2$H abundance and CS/SO ratio increase, as expected from previously published disk models \citep[e.g.,][]{2016ApJ...831..101B, 2018AA...617A..28S}. 

For physical model A, the SO emission between $\sim50-100$ au and the CS/SO ratio are best matched by the model with an input C/O ratio of 0.7, although they differ only modestly from the models with input C/O ratios of 0.47 and 0.9. An input C/O ratio of 1.0 significantly underpredicts SO emission at all radii. All of the models underpredict SO emission within a radius of 50 au and beyond a radius of 100 au. The models with input C/O ratios of 0.47 and 0.7 produce C$_2$H radial profiles with peak heights similar to the observed profile, but the model emission peaks at a larger radius compared to the observations. All of the models underpredict C$_2$H emission interior to $\sim$50 au and overpredict C$_2$H emission beyond $\sim100$ au. 

Compared to physical model A, physical model B yields fainter C$_2$H and SO emission relative to CS for a given input C/O ratio. Between $\sim50-100$ au, the model with an input C/O ratio of 0.47 best matches the observed SO emission. As with physical model A, physical model B underpredicts SO emission interior to a radius of $\sim50$ au and outside of $\sim100$ au for all input C/O values tested. All of the C$_2$H models significantly underpredict emission inside $\sim150$ au, but the overprediction of C$_2$H emission outside $\sim150$ au is less severe compared to physical model A. 

While decreasing the input C/O ratio would produce a better match to the observed SO emission in the inner 50 au, doing so would exacerbate the underprediction of C$_2$H emission. This suggests that the discrepancy in the SO and C$_2$H emission in this region is due to a deficiency either in the physical or chemical models. As shown in Figure~\ref{fig:physicalstructure}, the continuum and C$^{18}$O models slightly underpredict the emission in the inner disk compared to the observations, which suggests that an imperfect physical structure is at least partially responsible for the mismatch between the models and observations of the other lines. 

On the other hand, the underprediction of SO and overprediction of C$_2$H emission beyond 100 au nominally suggest that the models should have a lower gas-phase C/O ratio in this region. This is not readily achieved by simply lowering the input C/O ratio because one of the main oxygen carriers, H$_2$O, is largely frozen out in the outer disk. We ran tests with an input C/O value of 0.2 and found that the CS and SO column densities barely changed beyond 100 au. Along similar lines, the models in \citet{2021ApJS..257...12L} showed that the CS/SO values converged in the outer disk for input C/O values ranging from 0.5 to 1.5. 

\begin{figure*}
\centering
\includegraphics{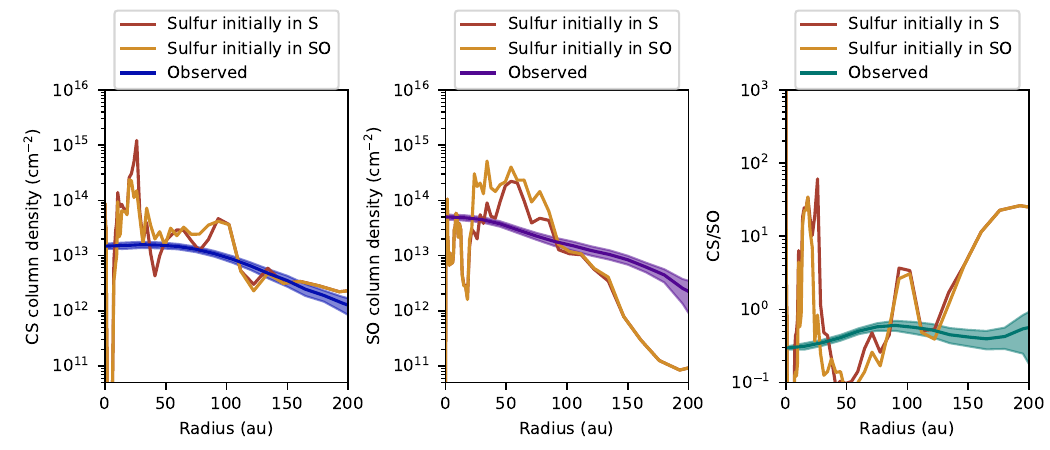}
\caption{A comparison of model results with sulfur starting in atomic S vs. SO. The input elemental sulfur abundance profile is the one derived for physical structure A and an input C/O ratio of 0.47, as shown in Figure \ref{fig:chemmodels}. The observed quantities correspond to the values from Figure~\ref{fig:CStoSOradprofiles} for an assumed excitation temperature of 30 K. \label{fig:testSOstart}}
\end{figure*}

We then examined whether the SO abundance in the outer disk can be better explained if an initial reservoir of SO is present. We took the input S abundance profile derived for an input C/O ratio of 0.47 and physical structure A (as shown in Figure \ref{fig:chemmodels}) and ran a new model under the extreme assumption that all sulfur is initially present in the form of SO. The input H$_2$O abundances were adjusted accordingly to maintain an overall C/O ratio of 0.47. A comparison of model results for sulfur starting in atomic S vs. SO is shown in Figure \ref{fig:testSOstart}. While starting sulfur in SO rather than atomic S boosts the SO column densities at radii less than 100 au, it makes little difference in the CS and SO column densities beyond 100 au. Thus, an initial reservoir of SO does not enhance the final abundance of SO in the outer disk because SO is efficiently converted to CS.  

\section{Discussion \label{sec:discussion}}
\subsection{Model limitations\label{sec:limitations}}

While production of CS and C$_2$H in disks is largely thought to proceed via gas-phase reactions, SO may have contributions from both gas-phase and grain surface chemistry \citep[e.g.,][]{2016ApJ...831..101B, 2018AA...617A..28S, 2021ApJS..257...12L}. The limited chemical network used in this study does not include grain surface pathways to produce SO, which may lead to underestimates of the C/O ratio required to reproduce the observed CS/SO ratio in DR Tau, particularly in the cold outer disk. However, models using comprehensive gas-grain networks have also found that CS/SO only falls below 1 in models for which C/O $<1$ \citep{ 2018AA...617A..28S, 2021ApJS..257...12L}, which supports the inference that C/O$<1$ in the regions of the DR Tau disk traced by our NOEMA observations. In addition, our C$_2$H observations serve as a separate check of C/O values that does not rely on grain surface chemistry. They indicate that C/O$<1$ at least beyond 100 au, although as noted in Section \ref{sec:models}, it is less clear from C$_2$H alone whether C/O$<1$ interior to 100 au because of uncertainties in the physical model.  

Another limitation of our models is that the vertical temperature and density structures are uncertain because the disk is nearly face-on and only one CO isotopologue line is used to constrain the models. The SED provides the primary constraint on DR Tau's vertical structure, which may be problematic if the extended material surrounding DR Tau's disk also contributes to the infrared emission. As in \citet{2022AA...660A.126S}, we find that relatively large values of the vertical settling parameter for large grains, $\chi\gtrapprox0.5$, are necessary to reproduce the SED's strong infrared emission. In other words, the large dust grains in our models are only moderately settled in comparison with the gas, yet observations of edge-on Class II disks indicate that they are often highly settled \citep{2020AA...642A.164V}. On the other hand, \citet{2023ApJ...946...70V} and \citet{2023ApJ...951....9L} found that the Class I disk IRAS 04302+2247, which has shed most of its envelope, exhibits only marginal levels of vertical settling. If the degree of settling increases with evolutionary stage, then DR Tau's $\chi$ values are plausible if it is a relatively young Class II disk or if the presence of the infalling material itself inhibits settling, as \citet{2023ApJ...946...70V} speculated. While both our models as well as the aforementioned models of other disks from the literature suggest that the gas-phase C/O is $<1$ in the outer regions of the DR Tau disk, the uncertainties in DR Tau's physical structure prevent us from placing tighter constraints. DR Tau's physical model can be improved by observing additional CO isotopologues spanning a range of upper state energy levels and optical depths, as illustrated by thermochemical modeling of other disks \citep[e.g.,][]{2021ApJ...908....8C, 2021ApJS..257...20S, 2022AA...663A..23L}. 

Our models do not include the effect of shock heating, which can enhance SO production \citep[e.g.,][]{1993MNRAS.262..915P,2021AA...653A.159V}. DR Tau features both spiral arms and infalling material, raising the possibility that either spiral shocks and/or accretion shocks could play a role in enhancing SO abundances in this system. Furthermore, while we neglected the spiral structure in our modeling, it is possible that some of the emission asymmetries observed in some of the other molecular species are related to the spiral structure. Higher resolution observations can help to determine whether there is an association.

\subsection{Comparison with other systems}

DR Tau's disk-averaged CS/SO of $\sim0.4-0.5$ is among the lowest values reported so far in the literature. In most cases, only lower bounds are available due to non-detections of SO. \citet{2018AA...617A..28S} estimated CS/SO $\gtrapprox1$ for the DM Tau disk, \citet{2021ApJS..257...12L} estimated CS/SO lower bounds ranging from $\gtrapprox4$ to $\gtrapprox14$ for the five disks from the MAPS ALMA Large Program, \citet{2021AJ....162...99F} estimated CS/SO $>109$ for the PDS 70 disk, and \citet{2023AA...675A.131T} estimated CS/SO $>1$ for the HD 142527 disk. Thus, DR Tau's disk-averaged CS/SO ratio ranges from at least a factor of 2 to 200 lower than these systems. \citet{2023AA...678A.146B} obtained a radially resolved estimate of CS/SO in the HD 169142 disk with values ranging from $\sim1-10$, also higher than that of DR Tau. A system with a significantly lower CS/SO value is the disk around the Herbig star Oph IRS 48, for which \citet{2021AA...651L...6B} estimated an upper bound of $<0.012$. While the Oph IRS 48 disk has the lowest CS/SO value reported for any disk thus far, DR Tau has the lowest CS/SO value reported for a T Tauri disk, which is relevant for understanding the possible formation conditions of planets around solar-mass stars. A major caveat in comparing the results of these studies, though, is that they did not observe the same transitions and most of them (including this work) had to make assumptions about the excitation temperature. A uniform, multi-transition survey of CS and SO in disks will be necessary to ascertain rigorously how CS/SO varies across the disk population. 

\begin{figure*}
\centering
\includegraphics{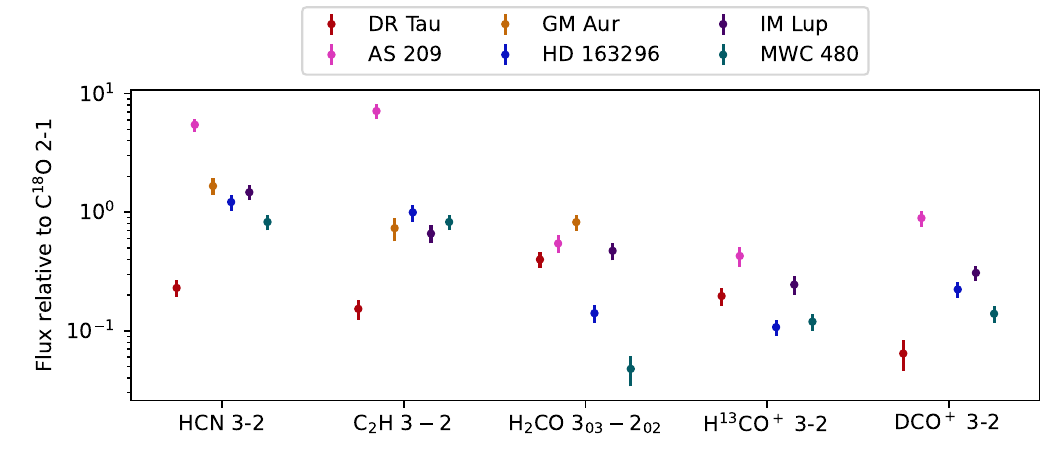}
\caption{Comparison of line flux ratios between DR Tau and the disks from the MAPS ALMA Large Program. \label{fig:fluxcomparisons}}
\end{figure*}

However, most of the other lines detected towards DR Tau are ones that have been commonly observed in disks, enabling a more direct comparisons of their chemical properties. Figure~\ref{fig:fluxcomparisons} compares the ratio of line fluxes to C$^{18}$O $J=2-1$ fluxes for DR Tau and the five Class II protoplanetary disks observed in the MAPS ALMA Large Program \citep{2021ApJS..257....1O}. The C/O values estimated from C$_2$H observations of these disks range from $\sim0.8-2$ \citep{2018ApJ...865..155C, 2021ApJS..257....7B}. For the MAPS sources, the C$^{18}$O $J=2-1$ fluxes are taken from \citet{2021ApJS..257....1O} and the HCN $J=3-2$, C$_2$H $N=3-2$, and H$_2$CO $3_{03}-2_{02}$ fluxes are taken from \citet{2021ApJS..257....6G}. The DCO$^+$ and H$^{13}$CO$^+$ measurements come from \citet{2017ApJ...835..231H}, which was not part of the MAPS program, but observed four of the same sources. The DR Tau fluxes come from this work and \citet{2023ApJ...943..107H}. The plotted C$_2$H fluxes correspond to the sum of the contributions from the four hyperfine components targeted in this work. For DR Tau, since C$_2$H emission is only detected in the stacked image, which is effectively an average of the four components, we estimated the total flux from all observed components by multiplying the stacked image flux by 4. A $10\%$ systematic flux calibration uncertainty was assumed for all measurements. Whereas DR Tau's H$_2$CO/C$^{18}$O and H$^{13}$CO$^+$/C$^{18}$O flux ratios are within the range of values of the MAPS disks, its DCO$^+$/C$^{18}$O ratio is somewhat lower and the HCN/C$^{18}$O and C$_2$H/C$^{18}$O ratios are markedly lower than the MAPS disks. Both C$_2$H and HCN fluxes tend to increase with the C/O ratio \citep[e.g.,][]{2016ApJ...831..101B, 2018ApJ...865..155C}, so the low values observed toward DR Tau qualitatively support the picture of DR Tau harboring a more oxygen-dominated gas-phase chemistry compared to most disks observed so far. A significant difference between the MAPS disks and DR Tau is that the former have large radial extents and deep, wide gaps \citep{2018ApJ...869L..41A, 2018ApJ...869...17L, 2020ApJ...891...48H}, while the latter is radially compact and lacking in prominent substructures, although visibility modeling indicates that DR Tau may have some narrow and shallow gaps \citep{2019ApJ...882...49L, 2020MNRAS.495.3209J}. \citet{2021AA...653L...9V} found tentative indications that compact disks tend to have weaker C$_2$H emission, which they hypothesized was a result of efficient radial drift helping to maintain higher oxygen abundances in the gas. 

\begin{figure}
\centering
\includegraphics{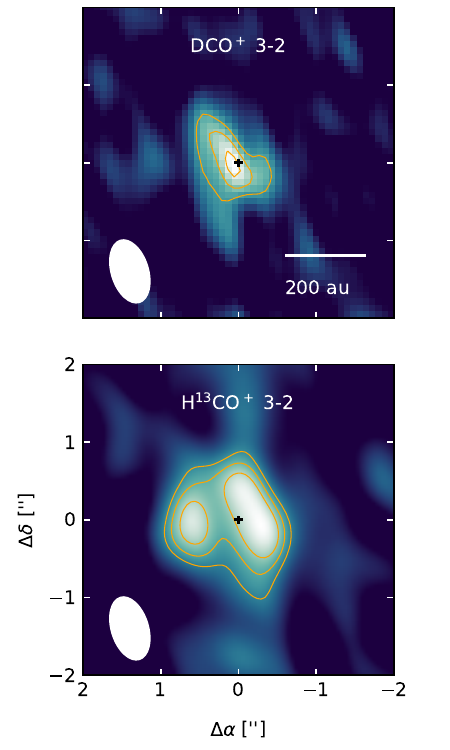}
\caption{A comparison of DR Tau's DCO$^+$ $3-2$ integrated intensity map from \citet{2023ApJ...943..107H} to the H$^{13}$CO$^+$ integrated intensity map from this work. The H$^{13}$CO$^+$ observations have been smoothed to match the resolution of the DCO$^+$ observations. Contours are drawn at the 3, 4, 5$\sigma$ levels.  \label{fig:DCOpcomparisons}}
\end{figure}

A comparison of DR Tau's DCO$^+$ and H$^{13}$CO$^+$ emission morphology (Figure~\ref{fig:DCOpcomparisons}) unveils a further peculiarity. The DCO$^+$ emission is comparatively compact and centrally peaked, while the H$^{13}$CO$^+$ emission is more radially extended and ring-like. This is the opposite of behavior seen in other disks, where DCO$^+$ emission tends to be more extended than H$^{13}$CO$^+$ emission \citep[e.g.,][]{2013AA...557A.132M, 2017ApJ...835..231H}. The characteristic relative distributions of DCO$^+$ and H$^{13}$CO$^+$ is ascribed to deuterium fractionation becoming more efficient in the colder regions of the outer disk \citep[e.g.,][]{1999ApJ...526..314A,2018ApJ...855..119A}. While it is possible that H$^{13}$CO$^+$ may be partially tracing DR Tau's envelope, this does not seem to account for the peculiar relative distributions of DCO$^+$ and H$^{13}$CO$^+$, since observations of embedded Class 0 and I sources indicate that DCO$^+$ should also be abundant in envelopes \citep[e.g.,][]{2021AA...655A..65T}. DR Tau's unusual emission patterns point to the possibility of a radial thermal inversion, which may also help to explain why it is able to maintain high levels of SO in its outer disk.  A couple of edge-on disks have shown evidence of radial thermal inversions in the form of CO emission enhancement in the midplane of the outer disk \citep{2017AA...607A.130D, 2021AJ....161..239F}. Such thermal inversions have been hypothesized to be due to stellar and/or interstellar radiation heating the midplane of the outer disk more readily as the dust optical depth drops \citep[e.g.,][]{2016ApJ...816L..21C, 2021AJ....161..239F}. Indeed, our physical model A for DR Tau features a modest midplane radial thermal inversion (see Figure~\ref{fig:physicalstructure}). The chemical networks used in this work do not include deuterated species, so examination of the impact of the thermal inversion on deuterium fractionation must be deferred to future studies. Alternatively, the presence of envelope material around DR Tau may keep disk temperatures elevated through effects such as backwarming \citep[e.g.,][]{1990ApJ...355..635K, 1994ApJ...420..326B} or accretion shocks \citep[e.g.,][]{1999ApJ...525..330Y}. Obtaining higher quality images of multiple transitions of DCO$^+$ and H$^{13}$CO$^+$ to measure their excitation temperatures and more robustly map their emission morphologies will be useful for determining whether DR Tau's disk indeed features a radial thermal inversion and what the likely origins of such an inversion are.

As noted in Section \ref{sec:models}, our overall sulfur abundances (S/H) estimated from modeling CS should be interpreted with caution since the SO emission profile is not reproduced at all radii. Nevertheless, it is useful to assess whether our values are reasonable through comparisons with other disks. Our model S/H values range from $\sim10^{-9}-10^{-5}$. \citet{2021ApJS..257...12L} tested models with spatially uniform abundances of $8\times10^{-8}$, $3.5\times10^{-6}$, and $1.5\times10^{-5}$ and found that an abundance of $8\times10^{-8}$ best matched the CS column densities toward MWC 480. \citet{2024MNRAS.528..388K} estimated a disk-averaged S/H value of $\sim5\times10^{-8}$ for HD 100546 from CS and SO observations, with local values rising as high as $\sim7\times10^{-7}$. As with our models of DR Tau, they were not able to fully reproduce CS and SO with the same input sulfur abundances. \citet{2018AA...617A..28S} found that an S/H value of $9\times10^{-7}$ best reproduces the CS column density measured toward DM Tau. While our sulfur abundances are generally in line with those estimated in other disks, our peak value of $\sim10^{-5}$ at a radius of $\sim50$ au for the models with low C/O values is notably higher than estimates from other disks. However, a couple factors may be driving the S abundance estimates to artificially high values in this region. First, CS is likely optically thick in the inner disk, as suggested by the discrepancy between the CS column densities from DALI and the values estimated using the optically thin LTE approximation (Figure \ref{fig:coldensitycomparisons}). Second, the spatial resolution is relatively coarse, so the radial SO abundance profile may not be as sharply peaked as the ones we estimated. Higher resolution observations, as well as observations of rarer isotopologues, are needed to derive the S abundance profile more robustly. 

Based on modeling of C$^{18}$O emission, we estimated a CO depletion factor of $\sim0.08-0.09$ for DR Tau. Observations of other Class II disks indicate that depletions of one to two orders of magnitude compared to ISM levels are typical \citep[e.g.,][]{2017AA...599A.113M, 2020ApJ...898...97B}. Meanwhile, among the younger, embedded Class I disks, some do not exhibit CO depletion \citep[e.g.][]{2020ApJ...901..166V, 2020ApJ...891L..17Z}, while others are depleted up to an order of magnitude \citep{2020ApJ...898...97B}. Thus, DR Tau's depletion factor appears to be within the ranges of both Class I and Class II disks, and it is not clear whether late infall plays a significant role in setting the CO abundance. However, as noted in Section \ref{sec:limitations}, our vertical structure is uncertain, which can make a large difference in CO depletion estimates \citep[e.g.,][]{2022ApJ...925...49R,2022RNAAS...6..176B}. Observations of other transitions of C$^{18}$O and optically thinner isotopologues will improve CO depletion estimates for DR Tau. 

\subsection{Origins of DR Tau's SO clump}
One of DR Tau's most striking and unusual features is the northeast clump of SO emission. With the benefit of higher angular resolution compared to previous observations of DR Tau from \citet{2023ApJ...943..107H}, we revisit and extend their discussion of what processes might be responsible for this asymmetric SO emission.

Azimuthal asymmetries in SO emission in disks increasingly appear to be common and have been attributed to a variety of origins. The SO asymmetries observed toward HD 100546 and HD 169142 have been hypothesized to be due to ice sublimation around a hot, embedded protoplanet \citep{2023AA...669A..53B, 2023ApJ...952L..19L}. No protoplanets have been detected in the DR Tau disk \citep{2022AA...658A..63M} and the SO clump does not coincide with a disk gap, so we consider the protoplanet explanation less likely for this system. \citet{2021AA...651L...6B} found that SO emission was strongly enhanced at a prominent dust trap traced by millimeter continuum observations in the Oph IRS 48 disk. However, no millimeter continuum emission is detected at the site of DR Tau's SO clump, which seems to rule out an association with dust traps. 

\begin{figure*}
\centering
\includegraphics{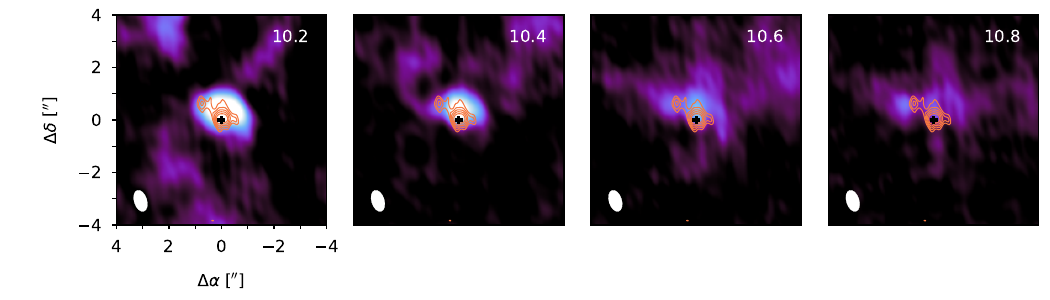}
\caption{$^{13}$CO $J=2-1$ channel maps \citep{2023ApJ...943..107H} showing the locations of streamer-like structures relative to the northeast SO clump. The orange contours correspond to the 4, 5, 6, 7, and 8$\sigma$ levels of the SO $7_6-6_5$ integrated intensity map. Black crosses mark the disk center. The LSRK velocity is provided in the upper right corner of each panel. \label{fig:streamercomparison}}
\end{figure*}

Observations of SO in embedded Class 0 and I systems may shed some light on the origins of the SO clump in the DR Tau disk. While only a handful of SO detections have been reported in Class II disks, it is commonly detected in younger sources \citep[e.g.,][]{2020ApJ...898..131L, 2023AA...678A.124A}. In a number of cases, SO emission appears to be enhanced at the envelope-disk interface or within infalling streamers \citep[e.g.,][]{2014Natur.507...78S, 2022AA...658A.104G, 2023ApJ...954..101A, 2023ApJ...953..190K, 2023ApJ...951...11Y}. This enhancement may be due either to the envelope/streamers delivering oxygen-rich material, or to localized heating from the infalling material that either sublimates oxygen-rich ices in the disk or promotes gas-phase production of SO \citep[e.g.,][]{2017ApJ...839...47M, 2021AA...653A.159V}. A comparison of DR Tau's $^{13}$CO $J=2-1$ channel maps (from \citealt{2023ApJ...943..107H}) to the SO $7_6-6_5$ integrated intensity map shows that there are several streamer-like structures traced by $^{13}$CO that may be connected to the SO clump (Figure~\ref{fig:streamercomparison}). However, since the $^{13}$CO observations are at a lower angular resolution than the SO observations, higher resolution $^{13}$CO observations will be required to ascertain definitively whether the SO clump coincides with a streamer. Low angular resolution ALMA ACA observations of [C I] also show possible streamer structures north of DR Tau \citep{2022AA...660A.126S}, so higher resolution observations of this species would also be useful for further examination of the possible relationship between streamers and localized SO enhancements.

As noted in Section~\ref{sec:molecularlines}, the CS/SO ratio appears to decrease at the clump compared to the rest of the disk. DR Tau is thus the second disk after HD 100546 \citep{2023NatAs...7..684K} (and the first T Tauri disk) to show evidence of an azimuthally varying CS/SO ratio (and by extension, an azimuthally varying C/O ratio). \citet{2023NatAs...7..684K} hypothesized that the azimuthal variations in CS/SO in the HD 100546 disk are due to shadowing by a protoplanet. Meanwhile, DR Tau's localized CS/SO variations appear to be linked to late infall, suggesting that diverse processes can modify a disk's CS/SO and C/O ratio.

\section{Summary\label{sec:summary}}
We obtained new NOEMA observations of CS, SO, and C$_2$H toward the DR Tau disk, which we used to constrain its gas-phase C/O ratio. Our findings are as follows:
\begin{enumerate}
\item Depending on the assumed excitation temperature, we estimate a disk-averaged CS/SO value of $\sim0.4-0.5$ for DR Tau, which is one of the lowest values reported for disks so far and the lowest value reported so far for a T Tauri disk. Previous CS/SO estimates for T Tauri disks have only been lower bounds due to non-detection of SO. 
\item Comparisons of the CS, SO, and C$_2$H emission to thermochemical models indicate that the gas-phase C/O ratio of DR Tau is $<1$, in contrast with most disks with reported C/O measurements. 
\item The SO integrated intensity maps feature a clump of emission at $\sim180$ au northeast of the star. This clump has no counterpart in other lines, and its CS/SO value is lower than that of the rest of the disk. Comparisons with $^{13}$CO observations suggest that the clump may be associated with streamer-like structures. 
\item We also report new detections of HCN, HCO$^+$, and H$^{13}$CO$^+$. Combined with other line observations of DR Tau, we find that its disk exhibits markedly different chemical properties from ALMA MAPS Large Program disks, further underscoring the chemical diversity of planet-forming environments.  
\end{enumerate}

Our analysis of DR Tau motivates a broader study of the chemistry of disks undergoing late infall. If DR Tau's comparatively low gas-phase C/O values are indeed linked to late infall, this would imply that planets that form in disks undergoing late infall could have significantly different atmospheric compositions from planets that form in isolated disks. In any case, our observations suggest that the factors governing the C/O ratio in disks are complex and that caution should be applied when using exoplanet atmosphere C/O ratios to infer their formation history, since disks exhibit a wide range of chemical behaviors. 

\vspace{1cm}
This work is based on observations carried out under project numbers W20BE and W21BE with the IRAM NOEMA Interferometer. IRAM is supported by INSU/CNRS (France), MPG (Germany) and IGN (Spain). We thank our NOEMA local contact, Ana Lopez-Sepulcre, for assistance with W20BE.   This paper makes use of the following ALMA data: ADS/JAO.ALMA\#2016.1.01164.S. ALMA is a partnership of ESO (representing its member states), NSF (USA) and NINS (Japan), together with NRC (Canada), MOST and ASIAA (Taiwan), and KASI (Republic of Korea), in cooperation with the Republic of Chile. The Joint ALMA Observatory is operated by ESO, AUI/NRAO and NAOJ. We also thank Arthur Bosman, Ryan Loomis, Myriam Benisty, and Leon Trapman for helpful discussions. We also thank the referee for comments improving the clarity of the manuscript. Support for J. H. was provided by NASA through the NASA Hubble Fellowship grant \#HST-HF2-51460.001-A awarded by the Space Telescope Science Institute, which is operated by the Association of Universities for Research in Astronomy, Inc., for NASA, under contract NAS5-26555. L. K. acknowledges funding via a Science and Technologies Facilities Council (STFC) studentship. J.A.S. is supported by the Dutch Research Council (NWO; grant VI.Veni.192.241).

\facilities{NOEMA, ALMA}

\software{\texttt{analysisUtils} \citep{2023zndo...7502160H}, 
\texttt{AstroPy} \citep{2013AA...558A..33A, 2018AJ....156..123A, 2022ApJ...935..167A}, \texttt{CASA} \citep{2022PASP..134k4501C}, \texttt{cmasher} \citep{cmasher}, \texttt{DALI} \citep{2012AA...541A..91B, 2013AA...559A..46B},
\texttt{eddy} \citep{2019JOSS....4.1220T}, \texttt{GILDAS} \citep{2013ascl.soft05010G}, \texttt{matplotlib} \citep{Hunter:2007}, \texttt{optool} \citep{2021ascl.soft04010D}, \texttt{scikit-image} \citep{scikit-image}, \texttt{SciPy} \citep{2020SciPy-NMeth}, \texttt{VISIBLE} \citep{2018AJ....155..182L}, \texttt{vis\_sample} \citep{2018AJ....155..182L}}

\appendix

\section{Channel Maps\label{sec:chanmaps}}

\begin{figure*}[!htp]
\begin{center}
\includegraphics{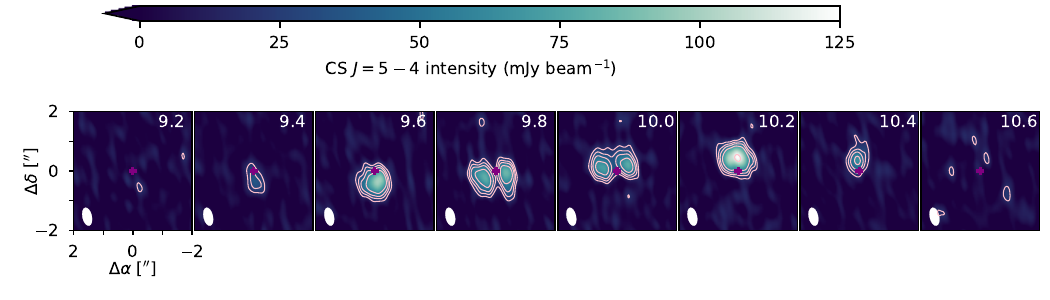}
\end{center}
\caption{Channel maps of CS $J=5-4$ toward DR Tau. The disk center is marked in each panel with a purple cross. Contours correspond to the [3, 5, 7, 10, 20]$\sigma$ levels. The top right corner of each panel shows the LSRK velocity in km s$^{-1}$, while the bottom left corner shows the synthesized beam. The complete figure set (7 images) is available in the online journal. \label{fig:chanmaps}}
\end{figure*}

Channel maps are presented in Figure~\ref{fig:chanmaps}.

\figsetstart
\figsetnum{13}
\figsettitle{DR Tau Channel Maps}
\figsetgrpstart
\figsetgrpnum{13.1}
\figsetgrptitle{CS $J=5-4$ channel maps}
\figsetplot{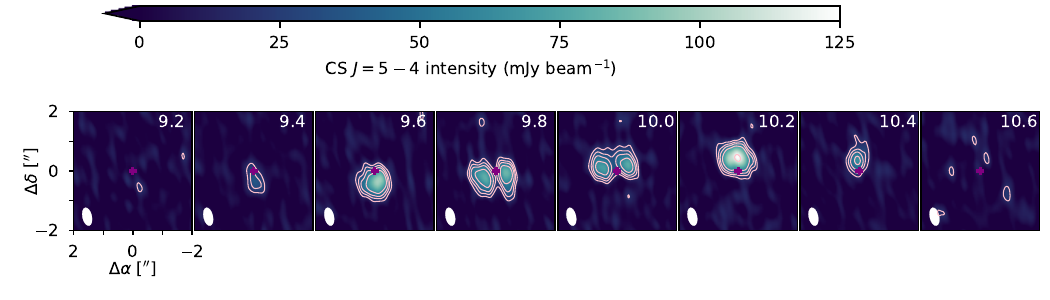}
\figsetgrpnote{Channel maps of CS $J=5-4$ toward DR Tau. The disk center is marked in each panel with a purple cross. Contours correspond to the [3, 5, 7, 10, 20]$\sigma$ levels. The top right corner of each panel shows the LSRK velocity in km s$^{-1}$, while the bottom left corner shows the synthesized beam.}
\figsetgrpend

\figsetgrpstart
\figsetgrpnum{13.2}
\figsetgrptitle{SO $7_6-6_5$ channel maps}
\figsetplot{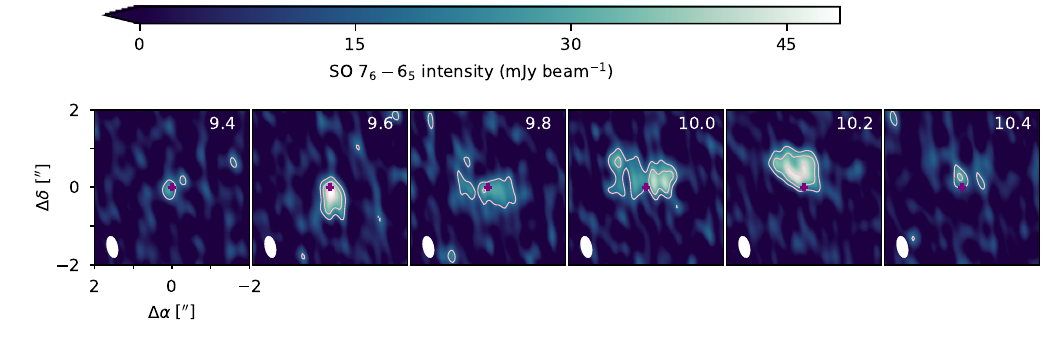}
\figsetgrpnote{Channel maps of SO $7_6-6_5$ toward DR Tau. The disk center is marked in each panel with a purple cross. Contours correspond to the [3, 5, 7]$\sigma$ levels. The top right corner of each panel shows the LSRK velocity in km s$^{-1}$, while the bottom left corner shows the synthesized beam.}
\figsetgrpend

\figsetgrpstart
\figsetgrpnum{13.3}
\figsetgrptitle{SO $5_6-4_5$ channel maps}
\figsetplot{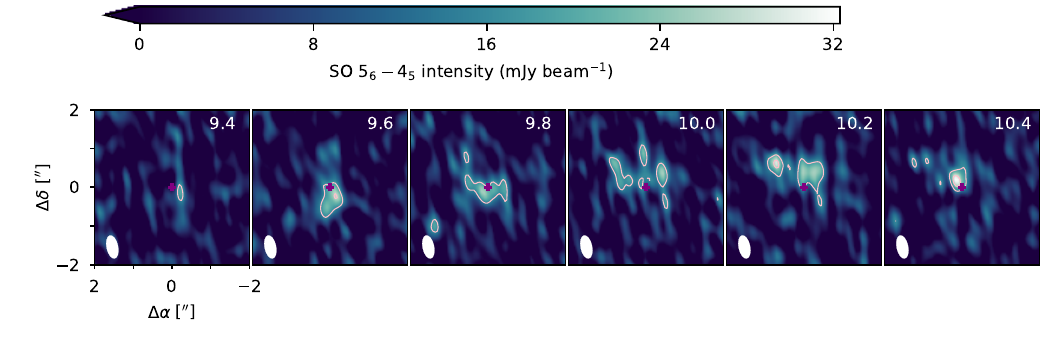}
\figsetgrpnote{Channel maps of SO $5_6-4_5$ toward DR Tau. The disk center is marked in each panel with a purple cross. Contours correspond to the [3, 5]$\sigma$ levels. The top right corner of each panel shows the LSRK velocity in km s$^{-1}$, while the bottom left corner shows the synthesized beam.}
\figsetgrpend

\figsetgrpstart
\figsetgrpnum{13.4}
\figsetgrptitle{C$_2$H $N=3-2$ stacked channel maps}
\figsetplot{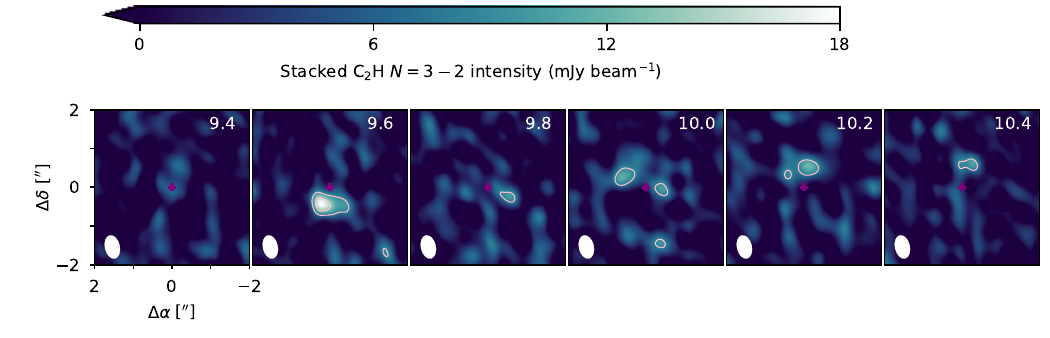}
\figsetgrpnote{Channel maps of the stacked hyperfine components of C$_2$H $N=3-2$ toward DR Tau. The disk center is marked in each panel with a purple cross. Contours correspond to the [3, 5]$\sigma$ levels. The top right corner of each panel shows the LSRK velocity in km s$^{-1}$, while the bottom left corner shows the synthesized beam.}
\figsetgrpend

\figsetgrpstart
\figsetgrpnum{13.5}
\figsetgrptitle{HCO$^+$ $3-2$ stacked channel maps}
\figsetplot{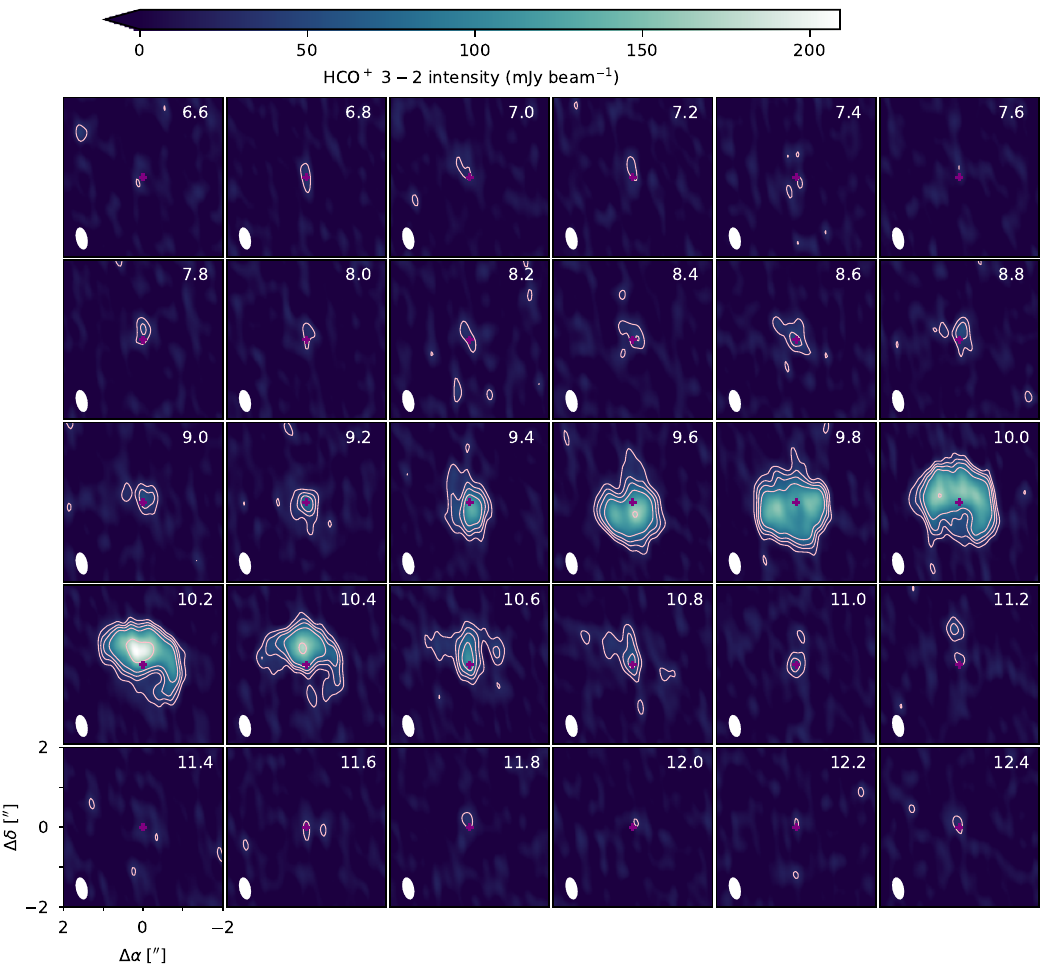}
\figsetgrpnote{Channel maps of HCO$^+$ $3-2$ toward DR Tau. The disk center is marked in each panel with a purple cross. Contours correspond to the [3, 5,7,10,20]$\sigma$ levels. The top right corner of each panel shows the LSRK velocity in km s$^{-1}$, while the bottom left corner shows the synthesized beam.}
\figsetgrpend

\figsetgrpstart
\figsetgrpnum{13.6}
\figsetgrptitle{H$^{13}$CO$^+$ $3-2$ stacked channel maps}
\figsetplot{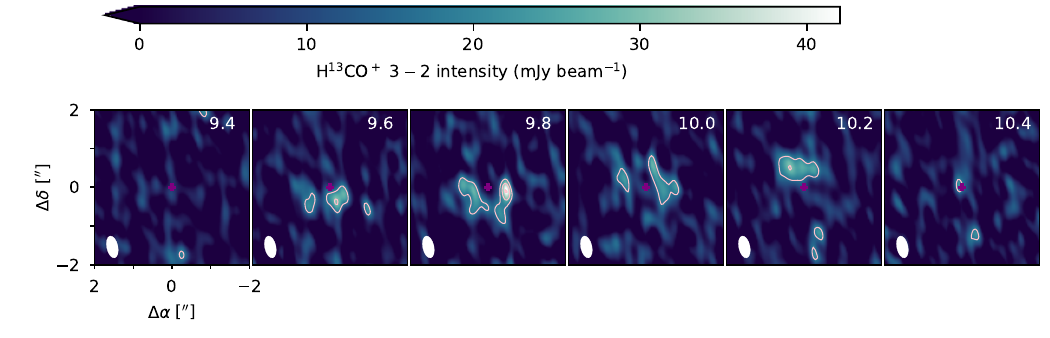}
\figsetgrpnote{Channel maps of H$^{13}$CO$^+$ $3-2$ toward DR Tau. The disk center is marked in each panel with a purple cross. Contours correspond to the [3, 5,7]$\sigma$ levels. The top right corner of each panel shows the LSRK velocity in km s$^{-1}$, while the bottom left corner shows the synthesized beam.}
\figsetgrpend

\figsetgrpstart
\figsetgrpnum{13.7}
\figsetgrptitle{HCN $3-2$ stacked channel maps}
\figsetplot{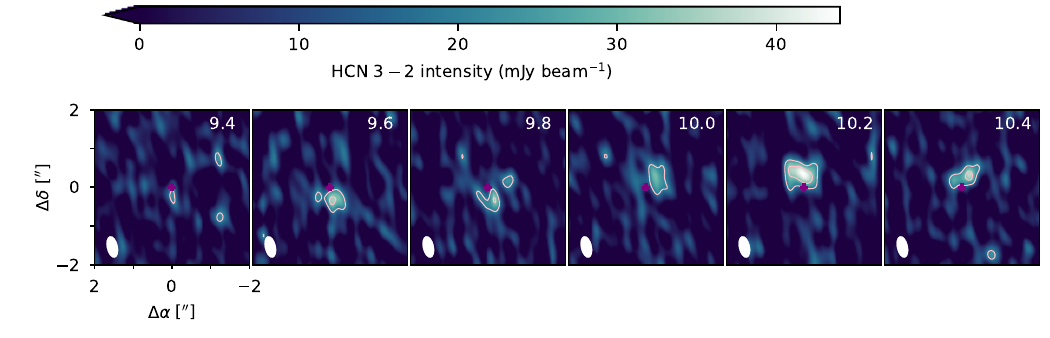}
\figsetgrpnote{Channel maps of HCN $3-2$ toward DR Tau. The disk center is marked in each panel with a purple cross. Contours correspond to the [3, 5]$\sigma$ levels. The top right corner of each panel shows the LSRK velocity in km s$^{-1}$, while the bottom left corner shows the synthesized beam.}
\figsetgrpend

\figsetend

\section{Details on modeling DR Tau's disk physical structure \label{sec:modeldescription}}
The DALI thermochemical code is described in detail by \citet{2012AA...541A..91B, 2013AA...559A..46B}. In brief, the key user inputs to DALI are the gas and dust density structure, stellar radiation field, chemical network, and initial chemical abundances. The standard version of DALI fixes the UV background spectrum to that of \citet{1978ApJS...36..595D}. DALI first performs a Monte Carlo radiative dust transfer calculation to estimate the dust temperature and then iteratively calculates the gas temperature, abundances, and heating/cooling rates. DALI can also use a previously calculated temperature structure to evolve only the chemistry. Following the temperature, abundance, and non-LTE excitation calculations, DALI can raytrace continuum images and spectral image cubes.  

\subsection{Disk structure parametrization}
Since the C$^{18}$O and millimeter continuum emission do not appear to have significant asymmetries, we employ an axisymmetric disk parametrization. While scattered light has revealed spiral structure \citep{2022AA...658A..63M}, it is not incorporated in our models because it does not appear to significantly affect gas or dust surface densities on the scale of the NOEMA beam. We comment on possible implications of the spiral structure for disk chemistry in Section~\ref{sec:discussion}. 

Based on \citet{2006ApJ...638..314D}, we assume that the dust consists of two sub-populations, each with a size distribution described by $n(a)\propto a^{-p}$. The ``small grain'' (sg) population has grain sizes ranging from $a_\text{min}=0.005$ $\mu$m to $a_\text{max}=1$ $\mu$m, while the ``large grain'' (lg) population ranges from $a_\text{min}=0.005$ $\mu$m to $a_\text{max}=1000$ $\mu$m. The value of $p$ is fixed to 3.5, corresponding to the MRN distribution from \citet{1977ApJ...217..425M}. 

Exterior to the dust sublimation radius $r_\text{subl}$, the surface density profiles of the dust sub-populations and the gas are parametrized as
\begin{equation}
\Sigma(r) = \Sigma_c \left(\frac{r}{r_c}\right)^{-\gamma}\exp\left[-\left(\frac{r}{r_c} \right)^q\right],
\end{equation}
where $r$ is in cylindrical coordinates. When $q=2-\gamma$, this expression becomes the commonly used form of the exponentially tapered power law disk surface density profile based on similarity solutions \citep{1974MNRAS.168..603L, 1998ApJ...495..385H, 2008ApJ...678.1119H}. We employed a more general parametrization because initial experiments with the common form produced continuum emission that tapered off too gradually at larger radii compared to the observations. In addition, \citet{2022AA...660A.126S} found that DR Tau's continuum and CO emission were not well-fit by models that assumed that the surface density of large grains was proportional to that of the gas. Therefore, unlike \citet{2022AA...660A.126S}, we allowed $r_c$, $\gamma$, and $q$ to differ for the gas and large grains. 

The disk gas mass $M_\text{gas}$ was left as a free parameter, and $M_\text{gas}/M_\text{dust}$ was fixed to the ISM value of 100. The overall dust mass fraction of the large grain population is described by the parameter $f_{lg}$. However, the mass fraction of the large grains at a given location in the disk is generally not equal to $f_{lg}$ because the small and large dust grain populations are not coupled. The values of $\Sigma_{c,\,gas}$ and $\Sigma_{c,\,lg}$ were then derived through numerical integration of the surface density profiles to find the scaling factors needed to match $M_\text{gas}$ and $f_{lg}\times M_\text{dust}$. The small grain population was assumed to be coupled to the gas, so the ratio of the small dust density to the gas density everywhere is $(1-f_{lg})/100$. 

The gas scale height is parametrized as 
\begin{equation}
H_{gas}(r) = H_{100}\left(\frac{r}{100\text{ au}} \right)^\psi.  
\end{equation}
To account for dust settling, the large grain scale height is parametrized as $H_{lg}(r) = \chi H_{gas}(r)$. The densities are then parametrized as 
\begin{equation}
\rho(r,z) = \frac{\Sigma_c}{\sqrt{2\pi}H(r)}\exp\left[-0.5\left(\frac{z}{H(r)}\right)^2\right]. 
\end{equation}

As in \citet{2022AA...660A.126S}, the stellar spectrum is modelled with two components. The first component accounts for the stellar photosphere and consists of a blackbody with a temperature of 3850 K and bolometric luminosity of 1.12 $L_\odot$, corresponding to the stellar values derived for DR Tau by \citet{2019AA...632A..32M}. Following \citet{2016AA...592A..83K}, the second component accounts for the UV excess associated with pre-main sequence stars and consists of a blackbody with a temperature of $10^4$ K and an accretion luminosity of 2.8 $L_\odot$, which corresponds to the DR Tau accretion luminosity derived by \citet{2019AA...632A..32M}. 

The PAH abundance was fixed to $0.1\times$ the ISM value, consistent with the value used by \citet{2022AA...660A.126S}. DALI defines ``ISM abundance'' to be 0.05\% of the gas mass. To our knowledge, the PAH abundance has not been constrained for DR Tau, but observations of other disks suggest that values of $0.1-0.01\times$ the ISM abundance are typical \citep{2006AA...459..545G}. \citet{2012AA...541A..91B} found that varying the PAH abundance in models primarily affected emission lines tracing the surface layers of the disk, while low-lying CO lines (like the transition we observe) did not significantly change. The adopted source properties are listed in Table~\ref{tab:sourceproperties}.

\begin{deluxetable}{lcc}
\tablecaption{Adopted source properties \label{tab:sourceproperties}}
\tablehead{
\colhead{Parameter}&\colhead{Value}&\colhead{Reference}}
\startdata
Distance & 192 pc & \citealt{2021AJ....161..147B}\\
$M_\ast$ & 1.2 $M_\odot$& \citealt{2021ApJ...908...46B}\\
$v_{sys}$ & 9.9 km s$^{-1}$& \citealt{2021ApJ...908...46B}\\
$i$ & $5.4^\circ$ & \citealt{2019ApJ...882...49L}\\
P. A. & $3.4^\circ$ &  
\citealt{2019ApJ...882...49L}\\
$T_{eff}$ & 3850 K & \citealt{2019AA...632A..32M} \\
$L_\ast$ & 1.12 $L_\odot$ &  \citealt{2019AA...632A..32M} \\
$L_{acc}$ & 2.8 $L_\odot$ &  
\citealt{2019AA...632A..32M} \\
$T_x$ & $3.51\times10^7$ K & \citealt{2019AA...625A..66D}\\
$L_x$ & $4.5\times10^{29}$ erg s$^{-1}$ & \citealt{2019AA...625A..66D} \\
Global gas-to-dust ratio & 100 & \\
$\zeta_{cr}$ & $10^{-18}$ s$^{-1}$& \\
PAH abundance relative to ISM & 0.1 & 
\enddata
\end{deluxetable}

\subsection{Dust modeling}
We adopted the DIANA standard dust opacities \citep{2016AA...586A.103W}, which were computed with the \texttt{optool} package \citep{2021ascl.soft04010D} using the distribution of hollow spheres method from \citet{2005AA...432..909M}. The forward scattering peak was ``chopped'' by two degrees (the \texttt{optool} default) to mitigate numerical issues that can arise with extreme forward scattering. 

DALI was used to produce model SEDs and 1.3 mm continuum images. Synthetic visibilities sampled at the same $uv$ points as the 1.3 mm continuum observations were generated from the model images using the \texttt{vis\_sample} package \citep{2018AJ....155..182L} and then CLEANed to compare with observations.

Disk modeling suffers from substantial degeneracies that are computationally expensive to explore \citep[e.g.,][]{2009ApJ...700.1502A,2023AA...672A..30K}. However, to obtain some idea of the degree to which our chemical inferences are sensitive to the adopted physical structure, we developed two sets of physical models, A and B. We fixed $f_{lg}$ to 0.99 for Physical Model A and to 0.90 for Physical Model B based on values commonly estimated or assumed for disks \citep[e.g.,][]{2016AA...592A..83K, 2021ApJS..257....5Z}. Previous astrochemical models have shown that molecular abundances in disks are sensitive to the abundance of small vs. large dust grains \citep[e.g.,][]{2006ApJ...642.1152A, 2019MNRAS.484.1563W, 2021ApJS..257....7B}. For both models A and B, the other disk structure parameters were initialized based on the best-fit DR Tau model from \citet{2022AA...660A.126S}. We then varied these values to improve the visual match between the dust models and the observations. As in \citet{2022AA...660A.126S}, all of our iterations produced markedly less emission at wavelengths shorter than 10 $\mu$m compared to the observations. DR Tau exhibits unusually strong emission shortward of 10 $\mu$m that has been hypothesized to be due to gas within the dust sublimation radius \citep{2011ApJ...730...73F}. We therefore focused only on reproducing the SED longward of 10 $\mu$m.  
\subsubsection{C$^{18}$O modeling}
The SED and millimeter continuum do not constrain $r_\text{c, gas}$ well. Therefore, once we identified preliminary disk structure parameter values that reasonably reproduced the SED and millimeter continuum radial intensity profile, we modelled C$^{18}$O in order to estimate $r_\text{c, gas}$.

\begin{deluxetable}{lc}
\tablecaption{Fiducial initial abundances\label{tab:initabundances}}
\tablehead{
\colhead{Species}&\colhead{Abundance relative to hydrogen nuclei}}
\startdata
H & 0.01   \\
H$_2$ & 0.495 \\
He & 0.14\\
CO\tablenotemark{a} & $1.35\times10^{-4}$\\
H$_2$O ice \tablenotemark{a} & $1.53\times10^{-4}$\\
CH$_4$ & $10^{-10}$ \\
N$_2$ & $3.75\times10^{-5}$\\
S \tablenotemark{a}& $10^{-8}$\\
Si$^+$ & $10^{-11}$\\
Fe$^+$ & $10^{-11}$\\
Mg$^{+}$ & $10^{-11}$\\
\enddata
\tablenotetext{a}{Values for these species are varied for different models.}
\end{deluxetable}

The fiducial initial abundances are listed in Table~\ref{tab:initabundances}. Our initial abundances are adopted from \citet{2021ApJS..257....7B} and \citet{2023AA...673A...7L}, except the N$_2$ initial abundance is taken from \citet{2018ApJ...865..155C}. Since previous observations have indicated that CO abundances in the warm molecular layer of disks can be significantly lower than ISM values \citep[e.g.,][]{2013ApJ...776L..38F, 2017AA...599A.113M, 2019ApJ...883...98Z}, we allow the initial CO abundance to vary by scaling the fiducial initial abundance with the depletion factor $f_\text{CO}$, where $f_\text{CO}$ can range between 0 and 1. A value of $f_\text{CO}=1$ represents no depletion relative to the ISM. The H$_2$O abundance is scaled accordingly to maintain a fiducial C/O ratio of 0.47, which corresponds to the median C/O of F, G, and K stars in the solar neighborhood \citep{2016ApJ...831...20B}. Since the CO abundances are manually scaled, the cosmic ray ionization rate was set to a relatively low value of 10$^{-18}$ s$^{-1}$ to avoid additional chemical reprocessing of CO, following the approach of \citet{2021ApJS..257....5Z} and \citet{2021ApJS..257....7B}. Thus, $f_\text{CO}$ corresponds to the depletion factor relative to ISM levels after accounting for freezeout and photodissociation. Alternatively, $f_\text{CO}$ corresponds to the depletion factor relative to ISM levels in the disk's warm molecular layer, where CO abundances are not affected by freezeout or photodissociation. 

As an aside, we note that the term ``CO depletion'' has been used in different ways in the literature, which may lead to some confusion when comparing studies. For example, \citet{2022ApJ...925...49R} and \citet{2023ApJ...953..183P} define CO depletion as a reduction in CO abundances from ISM levels beyond that accounted for by freezeout, photodissociation, and conversion of CO to other species. This would generally yield a more modest degree of depletion compared to the definition used in \citet{2021ApJS..257....5Z} and in this work, in which gas-phase CO that has been removed by processes such as chemical conversion is accounted for in the $f_\text{CO}$ factor rather than being directly modelled. We favor a definition of CO depletion that is referenced against ISM levels because it allows for a more straightforward comparison of disk properties inferred from different studies even if the modeling procedures are not the same. 

The X-ray plasma temperature $T_x$ was set to $3.51\times10^{7}$ K, corresponding to the high energy component fit to DR Tau's X-ray spectrum in \citet{2019AA...625A..66D}. The X-ray luminosity between 1 and 100 keV, $L_x$, was set to $4.5\times10^{29}$ erg s$^{-1}$, which was derived by scaling an isothermal bremsstrahlung spectrum \citep[e.g.,][]{2009ApJ...701..142G} to match the absorption-corrected X-ray flux measured towards DR Tau between 1 and 10 keV by \citet{2019AA...625A..66D}.

We then generated C$^{18}$O models with initial guesses for $f_\text{CO}$. The C$^{18}$O model runs each consisted of two steps. The first is a steady-state gas temperature calculation, starting from the dust temperature calculation corresponding to the input physical structure derived from modeling the SED and continuum. This calculation used the small chemical network from \citet{2013AA...559A..46B}, which in turn is adapted from UMIST06 \citep{2007AA...466.1197W}. The network contains 109 species and 1463 reactions, including two-body gas-phase reactions, freezeout, thermal and photodesorption, photodissociation and photoionization, successive hydrogenation of C, N, and O in ice, H$_2$ formation, cosmic-ray ionization and cosmic-ray induced FUV reactions, and charge exchange and recombination with PAHs/small grains. We then used the resulting temperature structure to perform a time-dependent chemistry-only run of DALI for 1 Myr, the approximate age of DR Tau \citep{2019AA...632A..32M}, using the ``ISO'' network from \citet{2016AA...594A..85M}. This network, which consists of 185 species and 5755 reactions, includes isotopologues for oxygen and carbon-bearing species and therefore enables us to account for the impact of selective photodissocation on C$^{18}$O abundances \citep[e.g.,][]{2014AA...572A..96M}. Given that this network is much larger, and the rare isotopologues do not meaningfully affect heating and cooling, we did not use the ``ISO'' network for the initial thermochemical calculation. We set the isotope abundance ratios to $^{12}$C/$^{13}$C = 69, $^{16}$O/$^{18}$O=557, and $^{18}$O/$^{17}$O=3.6, in accordance with local ISM values \citep{1999RPPh...62..143W}. For all DALI models, we employed the module from \citet{2017AA...605A..16F} that accounts for the grain size distribution in the chemistry calculations rather than assuming a uniform grain size, which is the default behavior. Following each run with the ``ISO" network, we raytraced C$^{18}$O $2-1$ with DALI, subtracted the continuum in the model image cubes, used $\texttt{vis\_sample}$ to generate synthetic visibilities sampled at the same $uv$ points as the C$^{18}$O observations, and then CLEANed these visibilities to compare with the observations. After comparing the model and observed C$^{18}$O radial intensity profiles, we adjusted the values of $r_{c,\,gas}$ and $f_\text{CO}$ and then regenerated the SED, 1.3 mm continuum, and C$^{18}$O models. This procedure was repeated until the model matched the observed C$^{18}$O radial intensity profile to within $\sim10\%$ interior to $r=200$ au. 

\section{DR Tau ALMA continuum data reduction\label{sec:DRTaucontinuum}}
1.3 mm continuum observations of DR Tau from program 2016.1.01164.S were retrieved from the ALMA archive. The data were originally presented in \citet{2019ApJ...882...49L}, which describes the observations in more detail. The raw data were re-calibrated with the ALMA pipeline in \texttt{CASA}. Channels with line emission were flagged, and the remaining channels were spectrally averaged to produce a set of continuum-only visibilities. We then applied three rounds of phase self-calibration to the continuum data using solution intervals of 60, 30, and 15 seconds, respectively, which improved the continuum S/N by a factor of 2. The final continuum image was produced with multiscale \texttt{CLEAN } \citep{2008ISTSP...2..793C} and a robust value of 0.5. The resulting beam was $0.13'' \times 0.10''$ $(39\fdg9)$. The flux measured within a circular aperture with a diameter of $1''$ is $126.4\pm0.5$ mJy, which is consistent with \citet{2019ApJ...882...49L}. The $1\sigma$ flux uncertainty was determined by randomly placing the same aperture in signal-free regions of the image and taking the standard deviation of 500 measurements.

\end{CJK*}
\end{document}